\documentclass{article}

\usepackage{arxiv}

\usepackage[utf8]{inputenc} 
\usepackage[T1]{fontenc}    
\usepackage{hyperref}       
\usepackage{url}            
\usepackage{booktabs}       
\usepackage{amsfonts}       
\usepackage{amsmath, amssymb}
\usepackage{nicefrac}       
\usepackage{microtype}      
\usepackage{cleveref}       
\usepackage{lipsum}         
\usepackage{graphicx}
\usepackage[numbers]{natbib}
\newtheorem{proposition}{Proposition}
\usepackage{doi}
\usepackage{subcaption}

\title{Involution game with migration and spatial heterogeneity of social resources}

\newif\ifuniqueAffiliation
\uniqueAffiliationtrue

\usepackage{authblk}

\newbox{\orcid}\sbox{\orcid}{\includegraphics[scale=0.009]{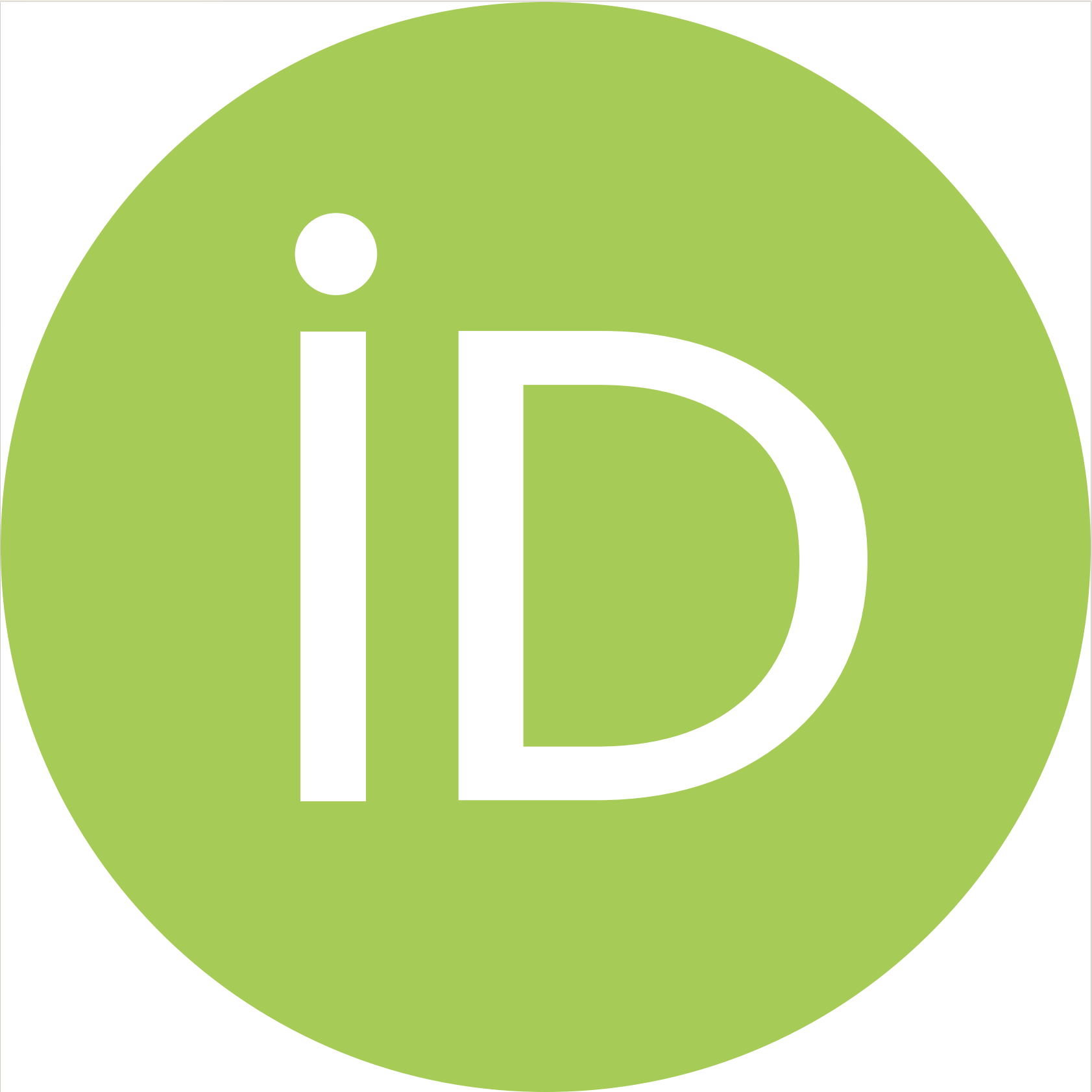}} 
\author[1]{%
	\href{https://orcid.org/0000-0001-9523-0818}{\usebox{\orcid}\hspace{1mm}Bo Li\thanks{\texttt{libo312@mails.ucas.ac.cn}}}%
}
\author[1]{%
	{Qiwen Ge\thanks{\texttt{597251335@qq.com, corresponding author.}}}%
}
\author[2,3,4]{%
	{Yong Shi\thanks{\texttt{yshi@ucas.ac.cn}}}%
}
\affil[1]{Wuhan City Polytechnic, Wuhan 430064, China}
\affil[2]{School of Economics and Management, University of Chinese Academy of Sciences, Beijing 100190, China}
\affil[3]{Research Center on Fictitious Economy and Data Science, Chinese Academy of Sciences, Beijing 100190, China}
\affil[4]{Key Laboratory of Big Data Mining and Knowledge Management, Chinese Academy of Sciences, Beijing, 100190, China}
\renewcommand{\shorttitle}{Involution game with migration and spatial heterogeneity of social resources}

\hypersetup{
pdftitle={Involution game with migration and spatial heterogeneity of social resources},
pdfsubject={q-bio.PE},
pdfauthor={Bo Li, Qiwen Ge and Yong Shi},
pdfkeywords={Involution, Migration, Resource heterogeneity, Evolutionary game, Agent-based model},
}

\begin{document}
\maketitle

\begin{abstract}
	Involution---a phenomenon of excessive competition with diminishing returns---has become a pressing socio-economic concern in contemporary China, prompting both academic inquiry and policy interventions.
	This paper proposes an evolutionary game model of involution that incorporates agent migration and spatial heterogeneity in resource distribution. 
	The model captures realistic features such as effort-based resource allocation, local interactions on a lattice, and mobility driven by payoff comparisons.
	We explore how varying conditions of migration and resource allocation influence the dynamics of involution. 
	The key findings from our simulations are as follows: when total resources are held constant, similar resource levels across different regions tend to suppress involution, whereas a large disparity between regions promotes it. Furthermore, increasing the total amount of resources exacerbates involution. 
	In addition, the probability of migration does not significantly affect the final evolutionary outcome. 
	We further identify threshold effects in the effort ratio and utility multiplier, revealing critical conditions under which involution emerges or subsides.
	To further elucidate these simulation results, we conduct a theoretical analysis using mean-field theory, which provides analytical expressions for the equilibria and stability conditions. 
	The theoretical predictions are in excellent qualitative agreement with simulation outcomes.
	Finally, we discuss real-world counterparts of the model, including competition among food delivery riders and between stores offering similar services.
\end{abstract}


\section{Introduction}\label{sec_intro}






The term ``involution'' actively discussed in the contemporary Chinese context refers to a phenomenon of excessive competitive behavior under conditions of limited resources, characterized by diminishing returns on individual efforts despite increased inputs \cite{TAO2025104531,WANG2022112092}. 
This socio-economic pattern has been observed across various domains, including labor markets, education, and commercial competition, where participants engage in escalating effort without proportional gains. 
Recently, in response to the recognized impediments posed by involutionary competition, China has implemented policy interventions aimed at mitigating its adverse effects on socio-economic advancement \cite{xiong2025anti}.
Understanding the mechanisms driving involution is therefore crucial for designing effective policies.

The involution dilemma arises when all agents would be better off adopting the low-effort strategy, yet each individual has an incentive to switch to the high-effort strategy (since it yields a higher personal payoff when others play low effort). However, when many choose high effort, the total resource is shared among greater total effort, leading to diminishing returns for everyone. Unlike the classic tragedy of the commons, the resource is not depleted --- effort only reallocates shares without creating new value. In short, the tragedy of the commons is about how much to \textbf{take} from a finite resource; the involution dilemma is about how much \textbf{effort} to invest for a fixed resource, where effort itself is costly and does not expand the resource pool.

At its core, involution shares structural similarities with classic social dilemmas such as the prisoner's dilemma \cite{nowak_evolutionary_1992, szabo_evolutionary_1997, szolnoki_impact_2009,   
	zhang_super-rational_2021} and the public goods game \cite{szolnoki_phase_2011, perc_evolutionary_2013, hauser_social_2019, latora_evolutionary_2021}---foundational concepts in game  theory that capture the tension between individual rationality and collective welfare.
This structural similarity has motivated researchers to employ evolutionary game theory---a well-established framework for analyzing strategic interactions---to examine the dynamics of involution, providing mechanistic insights into its emergence and potential pathways for mitigation \cite{WANG2022127307, li_involution_2023, huang_memory_based_2024, chaocheng_involution-cooperation-lying_2023}.

Building on these foundational insights, Wang et al. \cite{WANG2022112092} proposed a spatially explicit model on a square lattice where agents choose between two strategies distinguished by effort levels. 
Resources are allocated according to a maximum entropy principle, and agents adjust their strategies based on payoff comparisons with neighbors. 
Using the adoption ratio of the high-effort strategy as an indicator of involution intensity, they found 
that more abundant social resources can paradoxically intensify involution, while increased stochasticity in resource allocation suppresses it; moreover, the cost of high effort exhibits non-monotonic effects on involution dynamics.

In subsequent work, Wang et al. \cite{WANG2022127307} extended the model to incorporate spatio-temporal heterogeneity in social resources. 
They found that spatial heterogeneity has a dual effect on involution: it mitigates competition when resources are abundant (via network reciprocity), but intensifies it under moderate resource levels. 
Temporal heterogeneity, however, erases these spatial effects and restores homogeneous-population dynamics. 
Meanwhile, Li \cite{li_involution_2023} introduced a specialization strategy wherein agents concentrate efforts on specific resource niches, revealing that the adoption of specialization depends intricately on resource abundance, effort costs, and stochasticity.
Huang et al. \cite{huang_memory_based_2024} investigated memory impacts in spatial involution games, and Wang et al. \cite{chaocheng_involution-cooperation-lying_2023} proposed an evolutionary game model on a square lattice to abstract the fierce social competition between involution, cooperation, and lying flat.

While the above models focus specifically on involution, broader evolutionary game research has long illuminated how spatial structure and updating rules---especially migration---affect cooperation. 
Pioneering works date back to the early 1990s \cite{nowak_evolutionary_1992} and were systematically developed on regular lattices \cite{szabo_evolutionary_1997}. 
Seminal contributions that established the importance of spatial topology and migration include Szolnoki et al. \cite{szolnoki_topology-independent_2009}, Helbing and Yu \cite{doi:10.1073/pnas.0811503106}, and Chen et al. \cite{chen_risk-driven_2012}. 
These works demonstrated that migration toward successful individuals can promote cooperation under noisy conditions and that spatial heterogeneity fundamentally alters evolutionary outcomes. 
More recent studies \cite{WANG2022128097, ZHANG2022127073, li_effects_2021, shaw_gaps_2023, buesser_opportunistic_2013, xiao_environment-driven_2022, li_effect_2015, dhakal_evolution_2022, sun_adaptive_2026} have built upon this line of inquiry. 
For instance, the interplay between local and global strategy updating in public goods games significantly influences cooperative outcomes \cite{WANG2022128097}. 
More importantly for our purposes, migration based on environmental comparison promotes cooperation \cite{ZHANG2022127073}, highlighting the importance of mobility in strategic evolution. 

This observation motivates the central question of our study: how does agent mobility interact with spatial resource heterogeneity to shape involution dynamics?

Despite the advances in involution modeling, existing evolutionary game models have largely neglected agent mobility---a critical feature of many real-world competitive contexts. 
In reality, delivery workers, couriers, and homogeneous stores often compete for limited resources within regions while retaining the ability to relocate to areas with more favorable effort-to-reward ratios. 
This mobility can fundamentally alter competitive dynamics, by redistributing competitive pressure across space and enabling agents to escape locally saturated markets, yet remains unexamined in current involution models.

To address this gap, we propose an evolutionary game model of involution that incorporates agent migration and spatial heterogeneity of social resources. 
Our model examines how varying levels of mobility and resource distributions influence involutionary outcomes, with direct applications to competitions among food delivery riders and homogeneous stores. 
The key findings from our simulations are as follows: (i) when total resources are held constant, similar resource levels across different regions tend to suppress involution; (ii) by contrast, increasing total resources exacerbates involution, even under balanced spatial distribution; and (iii) the presence or absence of migration (rather than its precise rate) is the critical factor — once migration is allowed, the exact migration probability has negligible impact on the final evolutionary outcome.

To further understand these observations, we turn to mean-field theory---a standard approach in evolutionary game theory with foundational contributions dating back to the early 2000s \cite{szabo2005phase, Vukov20026, achdou_introduction_2020, sandholm_evolutionary_2020, antonov_mean-field_2021}. 
The analysis reveals that migration dynamics are governed by resource disparities, inevitably concentrating agents in resource-rich regions, while strategy evolution exhibits threshold behavior determined by average resource levels. 
Importantly, the theoretical predictions align closely with our simulation observations, confirming that the presence of migration — not its exact rate — and total resource abundance fundamentally modulate involution intensity.

The remainder of this paper is organized as follows: Section~\ref{sec:model} introduces the model formulation; Section~\ref{sec:results} presents simulation results and analysis; Section~\ref{sec:theory} provides theoretical analysis to uncover underlying mechanisms; and Section~\ref{sec:conclusion} offers discussion and concluding remarks.

\section{Model}\label{sec:model}

We consider an $N\times N$ two-dimensional lattice with periodic boundary conditions. Each lattice point $(x,y)$ may be occupied by an agent. The resource distribution is given by a matrix $\mathbf{M}$ where
\[
m_{ij} = 
\begin{cases}
	M_1 & \text{for } i \le N/2,\\[2pt]
	M_2 & \text{for } i > N/2,
\end{cases}
\]
so that the lattice is divided into two regions with different resource levels. The total number of agents is fixed at $N_{\text{agents}}$, with an initial density $\rho = N_{\text{agents}}/N^2$.

\subsection{Agent Strategies and Utility}
Agents compete for a fixed local resource $M(j)$ by investing effort. The share an agent receives is proportional to its effort relative to the total effort of all agents competing at that site. Switching from low effort (strategy C) to high effort (strategy D) increases an individual's share at the expense of others, but does not enlarge the total resource. When many agents simultaneously escalate effort, each individual's share may actually decrease because the total effort grows faster than the individual's effort---this mechanism captures the essence of involution.

Each agent can adopt one of two strategies:
\begin{itemize}
	\item \textbf{Strategy C (low effort)}: Effort level $e_C$, utility $u_C = e_C$.
	\item \textbf{Strategy D (high effort)}: Effort level $e_D$, utility $u_D = \beta e_D$, where $\beta$ is a multiplier that scales the utility of high effort relative to low effort.
\end{itemize}

The strategy of agent $i$ at time $t$ is denoted as $S(i,t) \in \{C, D\}$. The corresponding effort and utility are:
\begin{equation}
	o(i,t) = \begin{cases}
		e_C, & \text{if } S(i,t) = C, \\
		e_D, & \text{if } S(i,t) = D,
	\end{cases}
	\quad \text{and} \quad
	u(i,t) = \begin{cases}
		u_C, & \text{if } S(i,t) = C, \\
		u_D, & \text{if } S(i,t) = D.
	\end{cases}
	\label{eq:strategy_utility}
\end{equation}

\subsection{Payoff Calculation}
Each agent participates in resource allocation at its own location and at its four neighboring locations (von Neumann neighborhood with periodic boundaries). The agent engages in five independent competitions (one at each of these five cells). In each competition, the agent invests its full effort \(e_i\) (where \(i\) denotes the agent's strategy) and incurs the full cost \(e_i\). The resource at a cell is shared among the agents competing at that cell in proportion to their efforts. After all five competitions, the agent's payoff is the average net gain over these five competitions:
\begin{equation}
	\pi(i,t) = \frac{1}{5} \sum_{j\in\mathcal{N}(i)} \left( \frac{e_i}{\sum_{k\in\mathcal{N}(j)} e_k} M(j) - e_i \right),
	\label{eq:payoff_detailed}
\end{equation}
where \(\mathcal{N}(i)\) denotes the set consisting of agent \(i\)'s location and its four neighbors, and \(M(j)\) is the resource at location \(j\). The term \(\frac{e_i}{\sum_{k\in\mathcal{N}(j)} e_k} M(j)\) represents the share of resource \(M(j)\) obtained by agent \(i\) at cell \(j\), and \(e_i\) is the cost incurred in that competition. The factor \(1/5\) outside the sum yields the average payoff per competition.

\subsection{Strategy Update Rule}
At each discrete time step $t$, a pair of agents $(i, i')$ is randomly selected without replacement, where $i$ is the focal agent and $i'$ is the comparison agent. 
The probability that agent $i$ adopts the strategy of agent $i'$ at the next time step $t+1$ follows a pairwise comparison rule \cite{nowak_evolutionary_1992, szabo_evolutionary_1997, traulsen_stochastic_2006}, a standard imitation dynamics that has been widely used in evolutionary game theory since the 1990s and formalized in economic contexts by \cite{schlag_why_1998}. 
Specifically, we adopt the logit rule (often called the Fermi function):
\begin{equation}
	p\left[ S(i,t+1) = S(i',t) \right] = \frac{1}{1+\exp\{[\pi(i,t)-\pi(i',t)]/k\}},
	\label{eq:fermi}
\end{equation}
where $k$ is the selection intensity (noise level). This update rule embodies the tendency to imitate more successful agents. It is worth noting that the Fermi function is only one of several possible functional forms for the imitation probability; alternative forms (e.g., linear or step functions) have also been explored in the literature.

\subsection{Migration Rule}
After the strategy update, the same pair $(i, i')$ is used for a potential migration event. Agent $i$ may decide to move to a location adjacent to agent $i'$, based on a comparison of their payoffs at time $t$. Specifically, agent $i$ randomly selects one of the four neighboring cells of agent $i'$ (von Neumann neighborhood). If that cell is empty, agent $i$ migrates there at time $t+1$ with probability:
\begin{equation}
	p_{\text{migrate}} = \frac{\mu}{1+\exp\{[\pi(i,t)-\pi(i',t)]/k\}},
	\label{eq:migration}
\end{equation}
where $\mu \in [0,1]$ is the migration rate parameter controlling the baseline propensity to migrate. If the selected neighbor of $i'$ is occupied, the migration attempt fails and agent $i$ remains at its current location. This rule captures the tendency of agents to cluster near successful individuals, reflecting real-world behaviors such as settling near profitable locations.

\subsection{Simulation Parameters}
Table~\ref{tab:params} summarizes the parameters used in our simulations, along with their default values and ranges where applicable. All simulations are run for 3000 Monte Carlo steps, results are averaged over 50 independent runs, and error bars (standard deviation) are shown in the figures.

We set the agent density to \(\rho = 0.2\) as a baseline. This choice preserves sufficient empty space for migration (avoiding artificial constraint on mobility) and isolates the involution mechanism from overcrowding effects. 
Higher densities would limit migration opportunities and introduce congestion-dominated competition, which is not the focus of this study. 

To exclude finite-size artifacts, we verified that the main results remain unchanged when the lattice size is increased from \(50\times50\) to \(200\times200\) while keeping the agent density constant. Although migration can be long-range owing to the random selection of the target agent, the system shows no detectable finite-size dependence beyond \(L = 100\) for both the strategy composition \(F_D\) and the spatial distribution \(P_{M_1}\).

\begin{table}[h]
	\centering
	\caption{Simulation parameters and default values.}
	\label{tab:params}
	\begin{tabular}{@{}llc@{}}
		\toprule
		\textbf{Parameter} & \textbf{Description} & \textbf{Default value / Range} \\
		\midrule
		$N$ & Lattice size (grid dimension) & $100$ \\
		$N_{\text{agents}}$ & Total number of agents & $2000$ (varied in Sec.~\ref{subsec:agent_number}) \\
		$\rho$ & Initial agent density ($N_{\text{agents}}/N^2$) & $0.2$ \\
		$e_C$ & Effort level of strategy C & $0.1$ \\
		$e_D$ & Effort level of strategy D & $0.2$ \\
		$\beta$ & Utility multiplier for strategy D ($u_D = \beta e_D$) & $1$ (varied in Sec.~\ref{subsec:beta}) \\
		$u_C$ & Utility of strategy C ($= e_C$) & $0.1$ \\
		$u_D$ & Utility of strategy D ($= \beta e_D$) & $0.2$ (when $\beta=1$) \\
		$k$ & Selection intensity & $1$ (varied in Sec.~\ref{subsec:migration_selection}) \\
		$\mu$ & Migration rate & $0$ to $1$ (varied) \\
		$M_1, M_2$ & Resource levels in the two regions & $M_1+M_2$ fixed at $2.0$, $3.0$, or $5.0$; ratio varied \\
		\bottomrule
	\end{tabular}
\end{table}

\section{Results and Discussion}\label{sec:results}

We systematically investigate how resource heterogeneity and agent mobility shape involution dynamics. 
The main observable is $F_D$, the fraction of agents adopting the high\text{-}effort strategy D. 
We also monitor the spatial distribution $P_{M_1}$, the proportion of agents located in the $M_1$ region.

\subsection{Joint effects of migration and selection intensity}
\label{subsec:migration_selection}

\begin{figure}[h]
	\centering
	\includegraphics[width=0.9\textwidth]{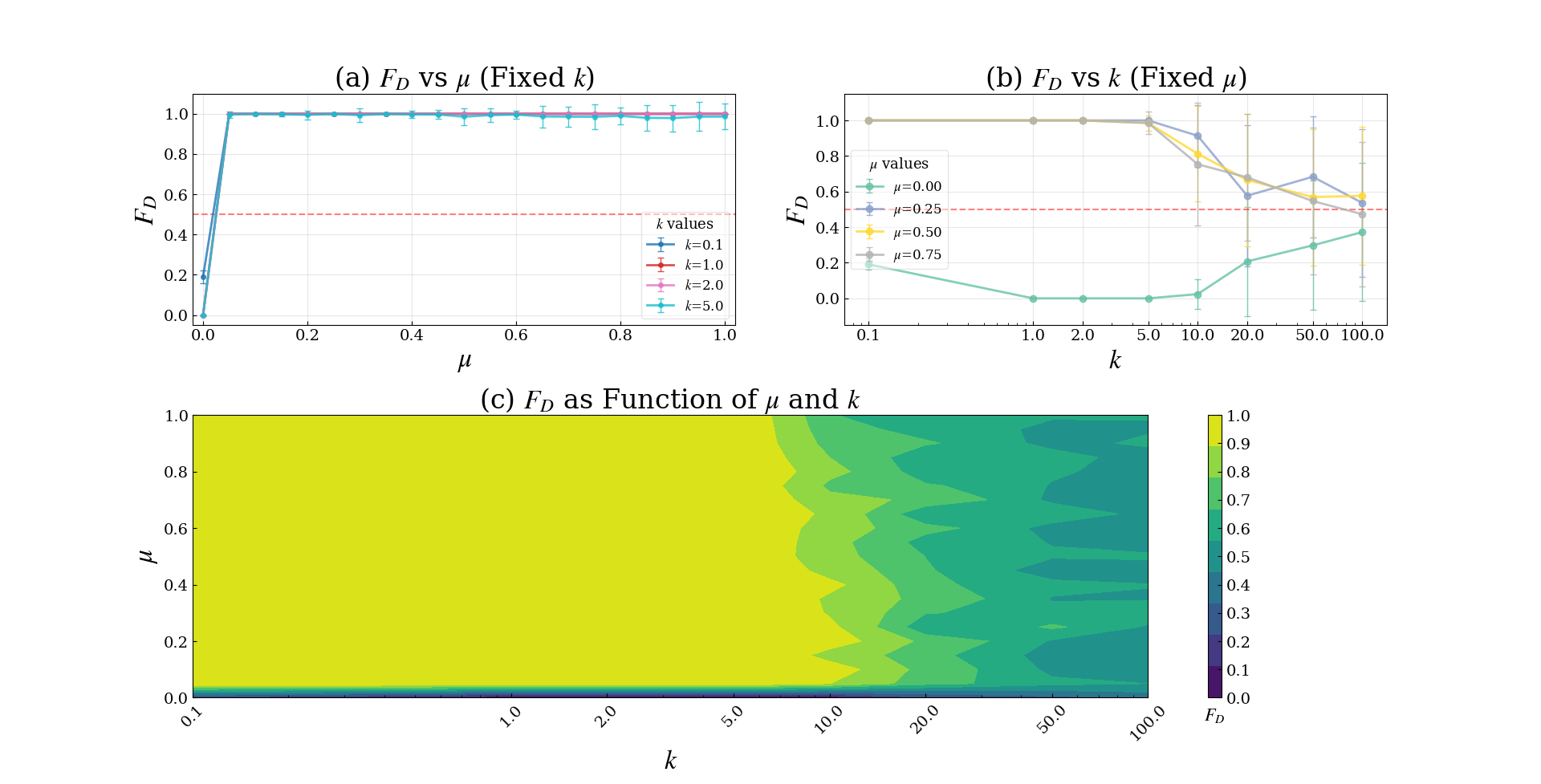}
	\caption{Proportion of competitive agents ($F_D$) as a function of migration rate $\mu$ and selection intensity $k$ under asymmetric resource distribution ($M_1 = 1$, $M_2 = 2$). 
		(a) $F_D$ vs.\ $\mu$ for different $k$; (b) $F_D$ vs.\ $k$ for different $\mu$; (c) contour plot over the whole parameter space. 
		Averaged over 50 independent runs on a $100\times100$ lattice with 2000 agents. For runs not reaching an absorbing state within 3000 steps, the last 300 steps are averaged. Error bars in (a) and (b) denote the standard deviation, where error bars are not visible, they are smaller than the symbol size. For the contour plot (c), when $k\leq5$, the standard deviation of 50 simulations is below 0.072. For large \(k\) (e.g., \(k \ge 50\)), the error bars become large (standard deviation up to \(\approx 0.4\)), indicating that the dynamics are dominated by stochastic fluctuations.}\label{fig1}
\end{figure}

We first analyze the proportion of competitive agents adopting the high-effort strategy ($F_D$) across varying migration rates ($\mu$) and selection intensities ($k$).
Figure \ref{fig1} elucidates the complex interplay between migration and selection intensity in determining involution levels under resource asymmetry ($M_1 = 1$, $M_2 = 2$). 
Panel (a) reveals that the presence of migration (any $\mu>0$) strongly suppresses competitive behavior compared to the immobile case ($\mu=0$) regardless of whether $k$ is 0.1, 1, 2, or 5, but once migration is allowed, the exact value of $\mu$ has little effect on $F_D$ across all selection intensities. 
Panel (b) further illustrates that, in addition to confirming the conclusions drawn from Panel (a) across a wider range of $k$ values, when $k$ exceeds a certain threshold (approximately $k = 5$), the final evolutionary outcome $F_D$ gradually converges toward $0.5$ as $k$ increases. 
The contour plot (c) synthesizes these trends, highlighting a low-migration high\text{-}involution regime, a high\text{-}migration low-involution regime, and a transitional region sensitive to both parameters.
Panel (c) synthesizes these trends in a comprehensive contour plot, highlighting three distinct regimes: (1) an immobile ($\mu=0$) and low $k$ regime characterized by low $F_D$ (low involution); (2) a mobile ($\mu>0$) and low-$k$ regime characterized by high $F_D$ (high involution); and (3) a regime at sufficiently large $k$ where $F_D$ gradually converges toward $0.5$ with increasing $k$, irrespective of the migration rate.
A more nuanced observation is that at intermediate selection intensities (when $k$ is approximately in the range of 10 to 30), increasing $\mu$ yields a modest additional suppression of the high-involution state, even after migration is introduced.
These findings suggest that the migration rate $\mu$ plays a moderating role in involution, the extent of which depends on the selection intensity $k$:
at low $k$, the presence of migration (any $\mu>0$) dramatically alters the evolutionary outcome compared to the immobile case, regardless of the specific migration rate; 
at intermediate $k$ (when $k$ is approximately in the range of 10 to 30), increasing $\mu$ further suppresses the high-involution state;
at sufficiently high $k$, the system converges toward a balanced equilibrium where $F_D$ approaches $0.5$, irrespective of the exact value of $\mu$.

However, as shown by the error bars in Fig.~\ref{fig1}, when the selection intensity is weak (large \(k\)), the system not only approaches a mixed equilibrium (\(F_D \approx 0.5\)) but also exhibits a high degree of run-to-run variability. For \(k \ge 50\), the standard deviation of \(F_D\) becomes substantial (typically \(0.3\)-\(0.4\)), regardless of the migration rate \(\mu\). This large variability indicates that under high noise levels, the evolutionary dynamics are dominated by stochastic fluctuations rather than deterministic payoff differences. In this regime, the Fermi function is nearly flat, so strategy updating effectively reduces to random drift. Consequently, different simulation runs may lock into different transient states, and the system fails to converge to a unique absorbing equilibrium within the finite simulation time. Our main conclusions in this paper focus on the more deterministic regime of moderate selection intensity (e.g., \(k = 1\)), where the error bars are small and the results are robust.
\subsection{Emergent spatial distribution of agents}
\label{subsec:spatial_distribution}

\begin{figure}[htbp]
	\centering
	\includegraphics[width=0.9\textwidth]{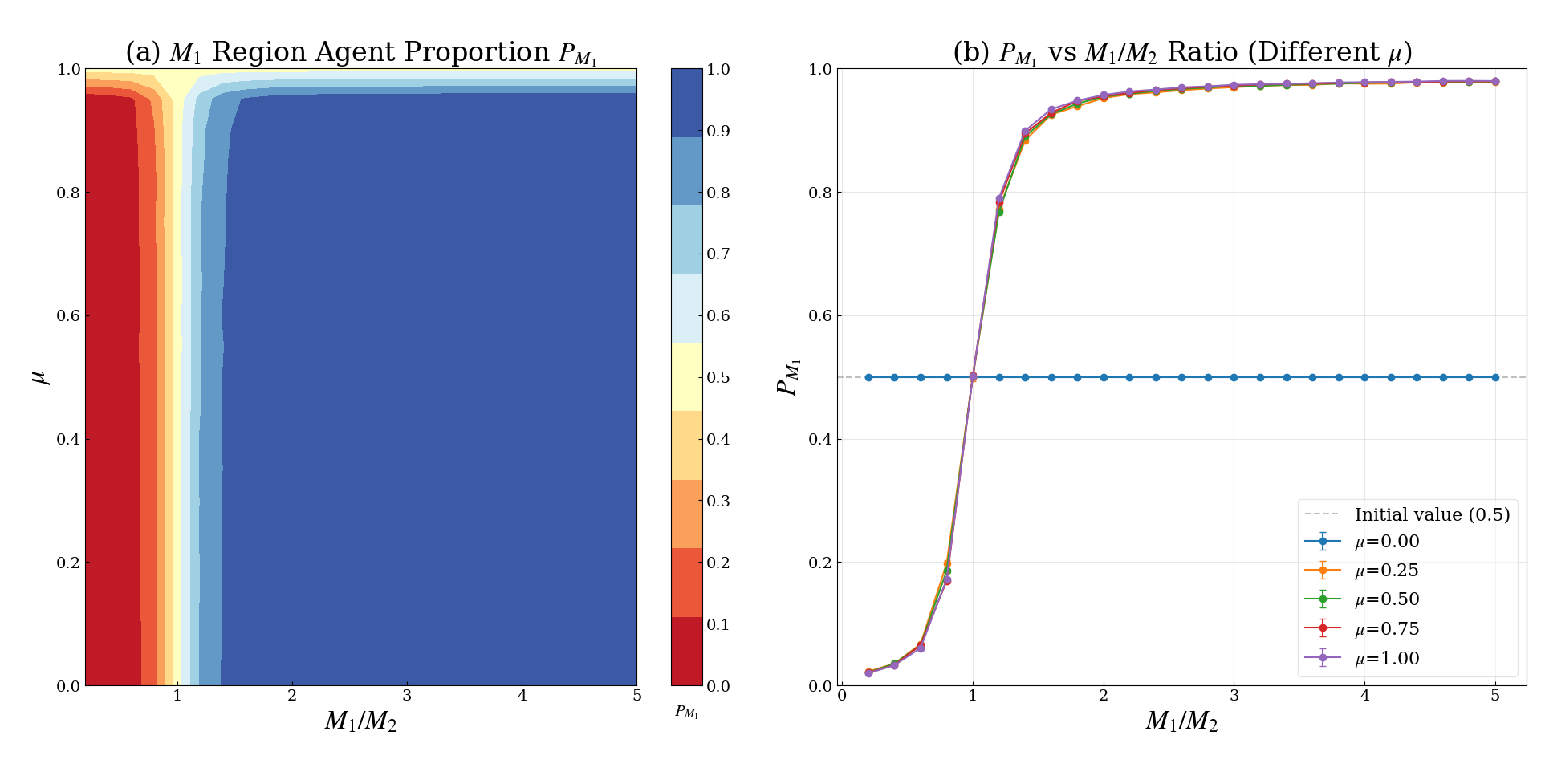}
	\caption{Proportion of agents located in region \(M_1\), \(P_{M_1}\), as a function of the resource ratio \(M_1/M_2\) for different migration rates \(\mu\). Simulations with total resources \(M_1+M_2 = 3.0\), 2000 agents on a \(100\times100\) lattice, averaged over 50 independent runs. The standard deviation of \(P_{M_1}\) across all simulations, including those shown in the panel (a), is below 0.05, making error bars invisible in panel (b). The dashed line at \(P_{M_1}=0.5\) indicates uniform distribution.}\label{fig:spatial_distribution}
\end{figure}

Figure~\ref{fig:spatial_distribution} examines how agents distribute themselves between the two regions in response to resource asymmetry and migration opportunities under a moderate total resource level (\(M_1+M_2=3.0\)). When migration is absent (\(\mu=0\)), agents remain uniformly distributed (\(P_{M_1}=0.5\)) for all \(M_1/M_2\), confirming that without mobility, spatial sorting does not occur. Once migration is allowed, agents rapidly concentrate in the resource-richer region. At \(\mu=1.0\), a modest resource advantage of \(M_1/M_2=1.5\) drives \(P_{M_1}\) to approximately \(0.90\), while at \(M_1/M_2=2.0\), \(P_{M_1}\) reaches \(0.96\). The concentration intensifies with both \(\mu\) and the degree of asymmetry, approaching complete agglomeration (\(P_{M_1}\to 1\) or \(0\)) for extreme ratios. The influence of \(\mu\) is particularly evident at intermediate ratios: for example, at \(M_1/M_2=1.2\), increasing \(\mu\) from \(0.05\) to \(1.0\) raises \(P_{M_1}\) from \(0.71\) to \(0.81\). These results demonstrate that migration consistently drives agents toward resource-rich areas, with the degree of concentration increasing with both resource disparity and migration rate. Such spatial sorting has direct implications for involution: concentration in resource-rich regions creates localized competition hotspots that may sustain high-effort strategies, highlighting the multi-scale nature of involution dynamics.

\subsection{Moderating role of resource abundance}
\label{subsec:resource_abundance}

\begin{figure}[h]
	\centering
	\includegraphics[width=6.7in]{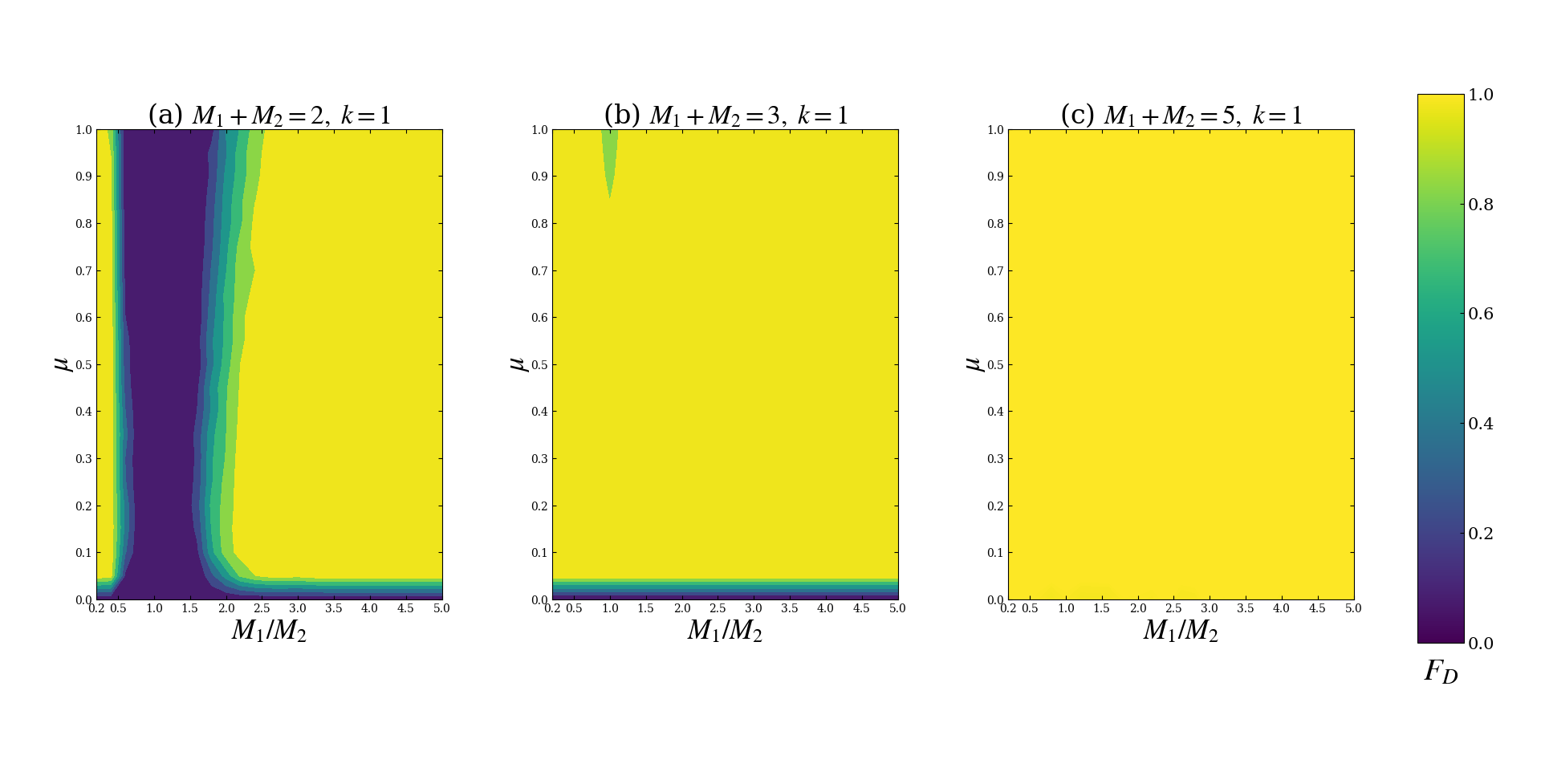}
	\caption{Proportion of competitive agents $F_D$ as a function of the resource ratio $M_1/M_2$ and migration rate $\mu$ under different total resource endowments. Each panel corresponds to a fixed total resource $M_1+M_2$ (2, 3, and 5, from left to right), with selection intensity fixed at $k=1$. Color intensity represents the fraction of agents adopting the high-effort strategy D. Results are averaged over 50 independent runs on a $100\times100$ lattice with 2000 agents. The standard deviation of $F_D$ across all simulation results is below 0.05.}\label{fig2}
\end{figure}

Figure~\ref{fig2} delineates the joint influence of total resource availability and spatial inequality on the prevalence of high-effort strategies. 
Three distinct regimes emerge as total resources increase from scarcity to abundance.

When total resources are severely limited ($M_1+M_2 = 2.0$), the system exhibits a sharp contrast between the immobile and mobile cases. 
In the absence of migration ($\mu = 0$), $F_D$ remains uniformly zero across all resource ratios, indicating that without mobility, agents never adopt the high-effort strategy under scarcity. Once migration is permitted ($\mu > 0$), however, $F_D$ becomes highly sensitive to the degree of resource inequality. 
For moderate ratios ($M_1/M_2$ between 2 and approximately 5), $F_D$ remains low, typically below 0.2, suggesting that mild asymmetry does not trigger widespread involution. 
As the ratio increases beyond 5-10, $F_D$ rises steadily, approaching 1 for extreme ratios ($M_1/M_2 > 15$). 
Notably, for any fixed $M_1/M_2$, varying $\mu$ has negligible impact on $F_D$; the critical distinction is simply whether migration exists or not. 
This pattern implies that under resource scarcity, mobility enables agents to concentrate in richer areas, but the resulting involution intensity is governed primarily by the magnitude of resource disparity rather than by the migration rate itself.

As total resources increase to a moderate level ($M_1+M_2 = 3.0$), the system undergoes a qualitative shift. 
At $\mu = 0$, $F_D$ is again zero across all ratios, reaffirming that without migration, involution does not arise. 
However, for any $\mu > 0$, $F_D$ jumps to near-unity for almost the entire parameter space. 
The only deviations occur at the lowest ratios ($M_1/M_2$ around 1-1.5), where $F_D$ dips slightly to approximately 0.8-0.9. 
This suggests that once resources reach a certain threshold, the presence of migration alone suffices to drive nearly universal adoption of the high-effort strategy, regardless of the precise degree of inequality. 
The migration rate $\mu$ again plays a secondary role, with its exact value having little effect on the final outcome.

When total resources are abundant ($M_1+M_2 = 5.0$), involution becomes essentially inevitable. 
Here, $F_D$ is uniformly close to 1 across all combinations of $\mu$ and $M_1/M_2$, including at $\mu = 0$. 
Even without migration, resource abundance alone is sufficient to push the entire population toward the high-effort strategy. Mobility, whether present or not, exerts no discernible influence on the equilibrium.

Taken together, these results reveal a hierarchical control of involution by total resource availability. 
Scarcity permits involution only under extreme inequality and requires migration to manifest; 
moderate abundance makes involution pervasive once migration is possible; 
abundance renders involution unconditional, independent of both mobility and spatial distribution. 
Across all regimes, the migration rate $\mu$ itself has minimal impact on $F_D$ only the presence or absence of migration matters. 
These findings underscore that policy interventions aimed at mitigating involution should prioritize addressing total resource levels and spatial disparities over merely facilitating mobility.

\subsection{Role of utility multiplier $\beta$}
\label{subsec:beta}

\begin{figure}[htbp]
	\centering
	\begin{subfigure}[b]{0.49\textwidth}
		\centering
		\includegraphics[width=\textwidth]{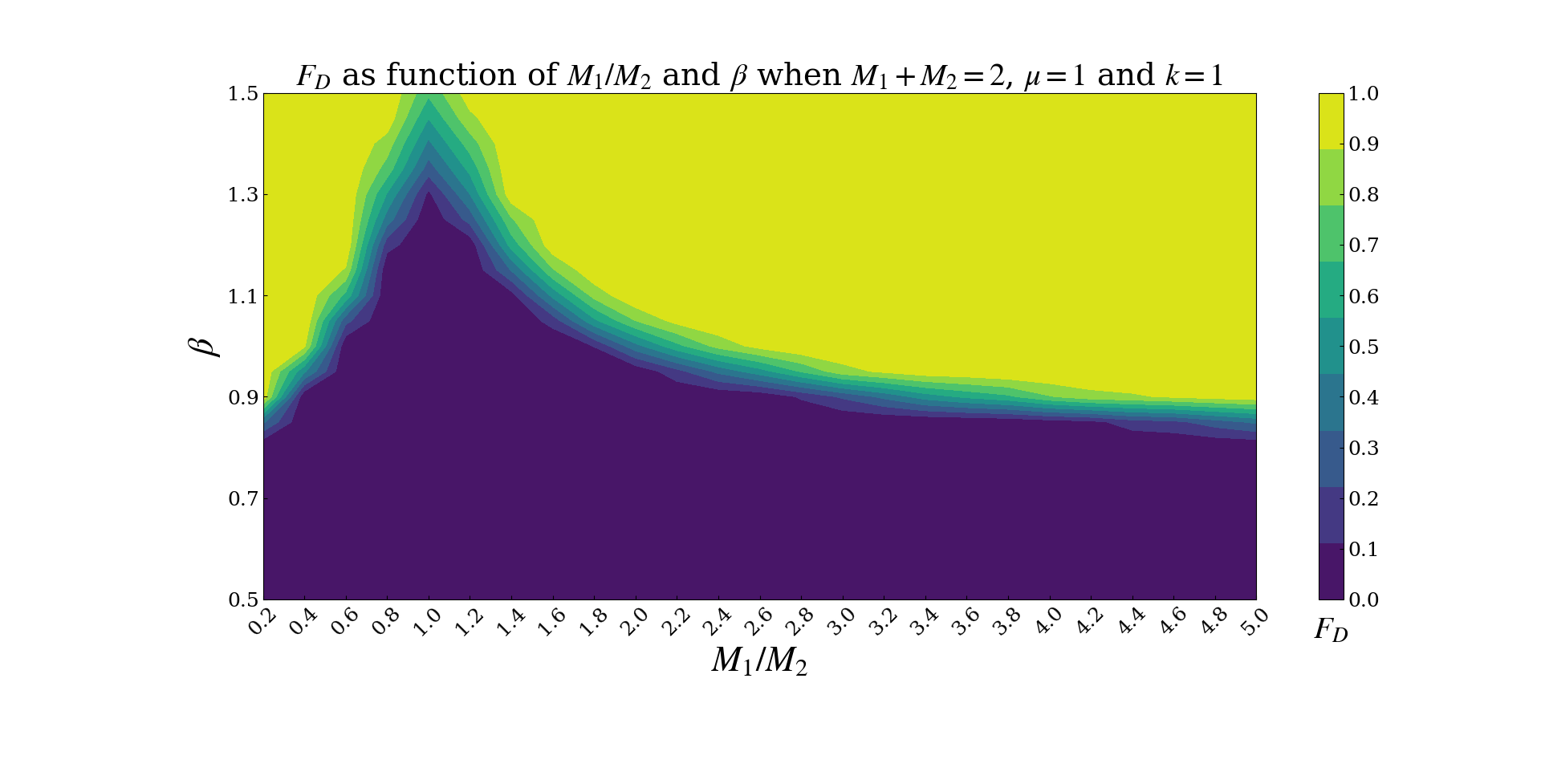}
		\caption{\(F_D\), \(M_1+M_2=2.0\)}
		\label{fig:beta_02a}
	\end{subfigure}
	\hfill
	\begin{subfigure}[b]{0.49\textwidth}
		\centering
		\includegraphics[width=\textwidth]{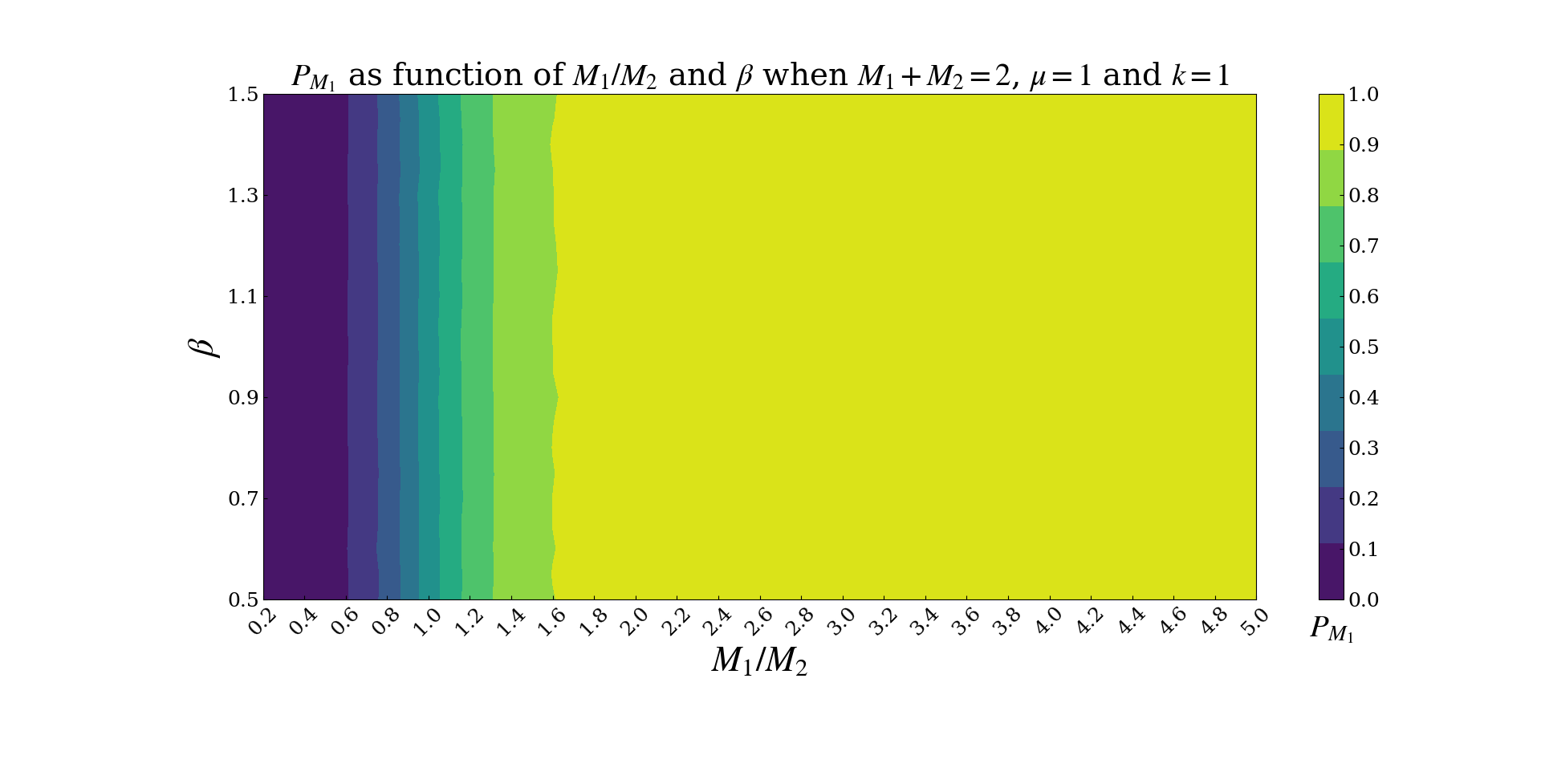}
		\caption{\(P_{M_1}\), \(M_1+M_2=2.0\)}
		\label{fig:beta_02b}
	\end{subfigure}
	
	\vspace{\baselineskip}
	
	\begin{subfigure}[b]{0.49\textwidth}
		\centering
		\includegraphics[width=\textwidth]{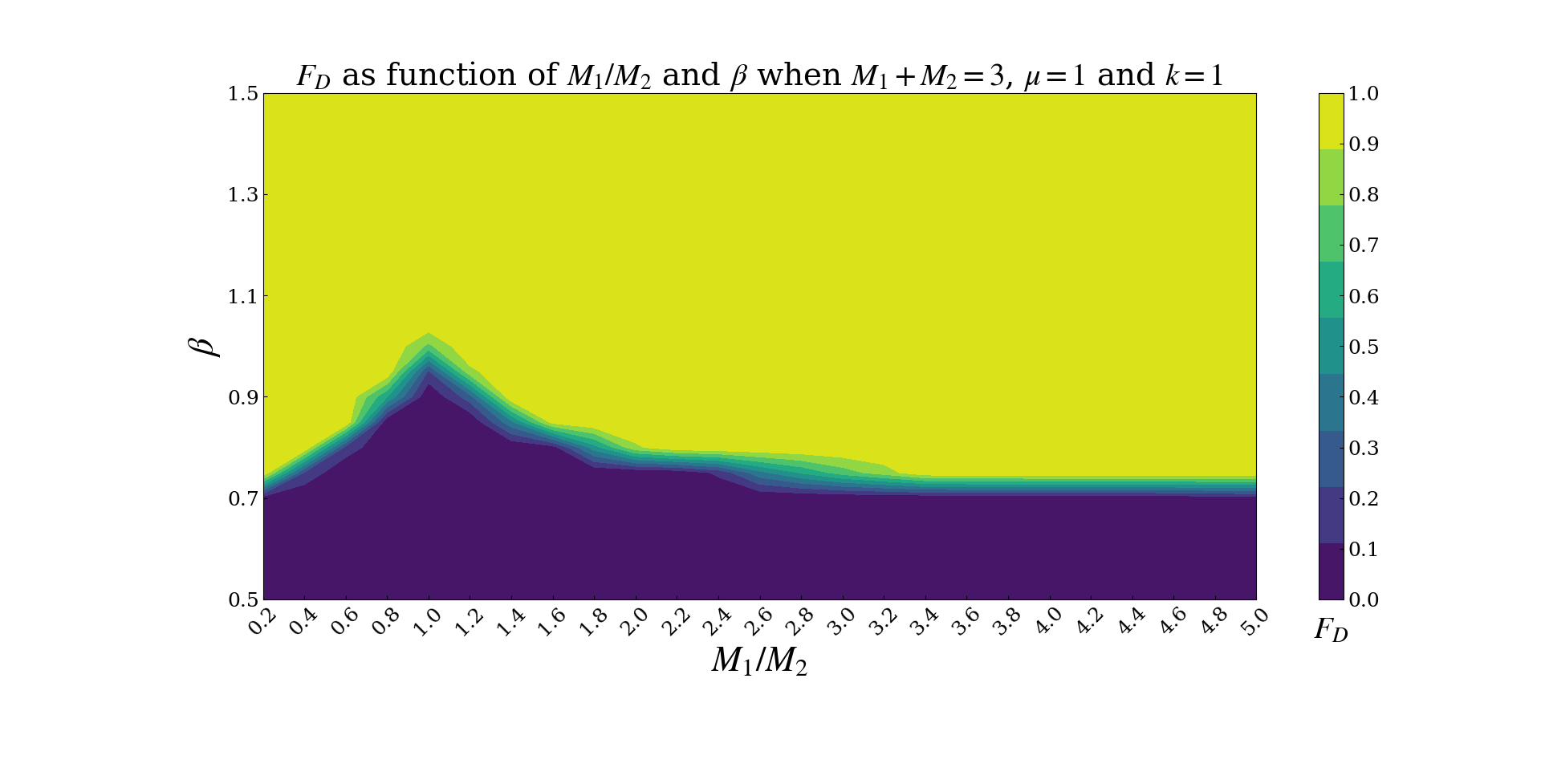}
		\caption{\(F_D\), \(M_1+M_2=3.0\)}
		\label{fig:beta_03a}
	\end{subfigure}
	\hfill
	\begin{subfigure}[b]{0.49\textwidth}
		\centering
		\includegraphics[width=\textwidth]{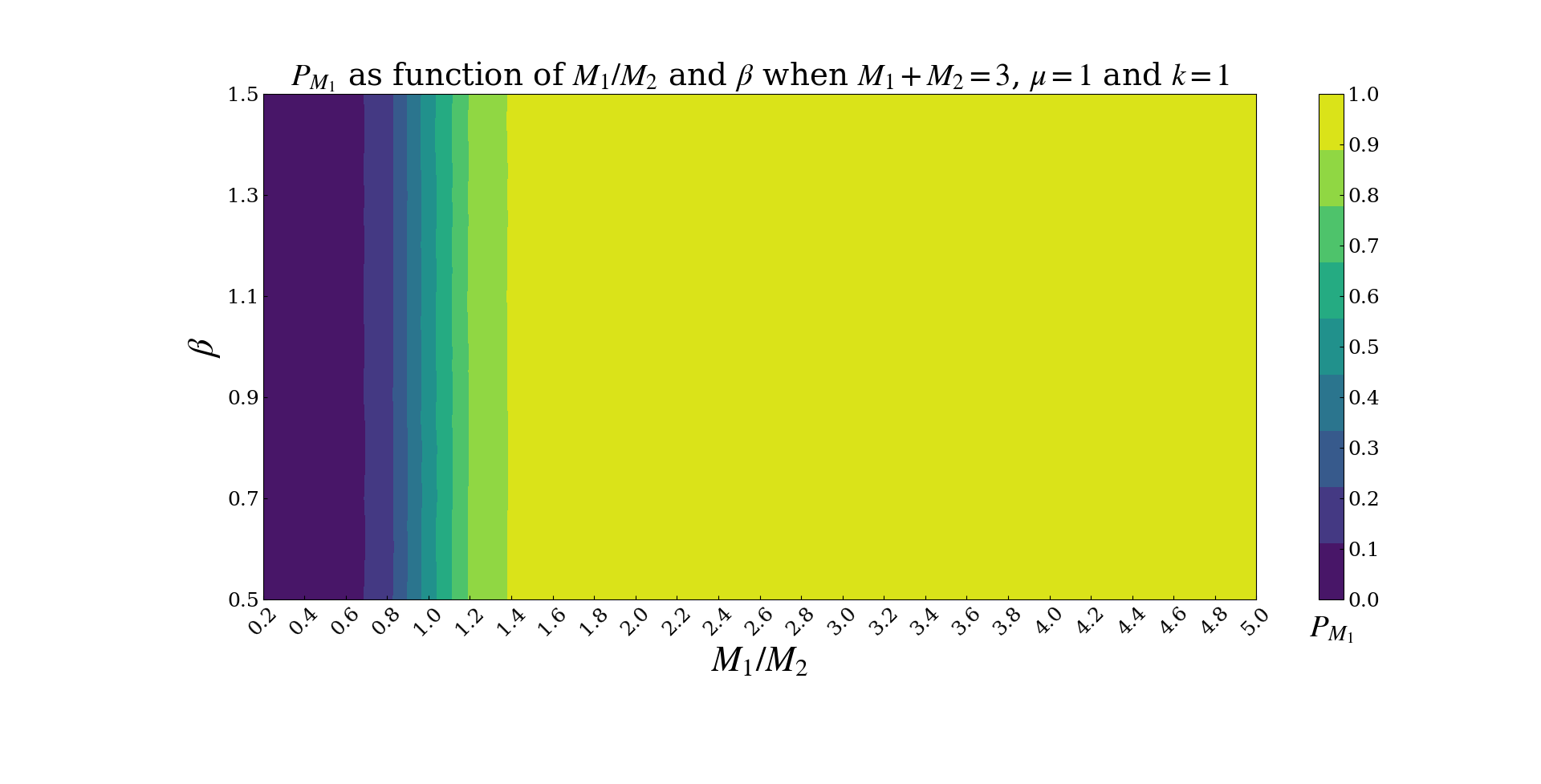}
		\caption{\(P_{M_1}\), \(M_1+M_2=3.0\)}
		\label{fig:beta_03b}
	\end{subfigure}
	
	\vspace{\baselineskip}
	
	\begin{subfigure}[b]{0.48\textwidth}
		\centering
		\includegraphics[width=\textwidth]{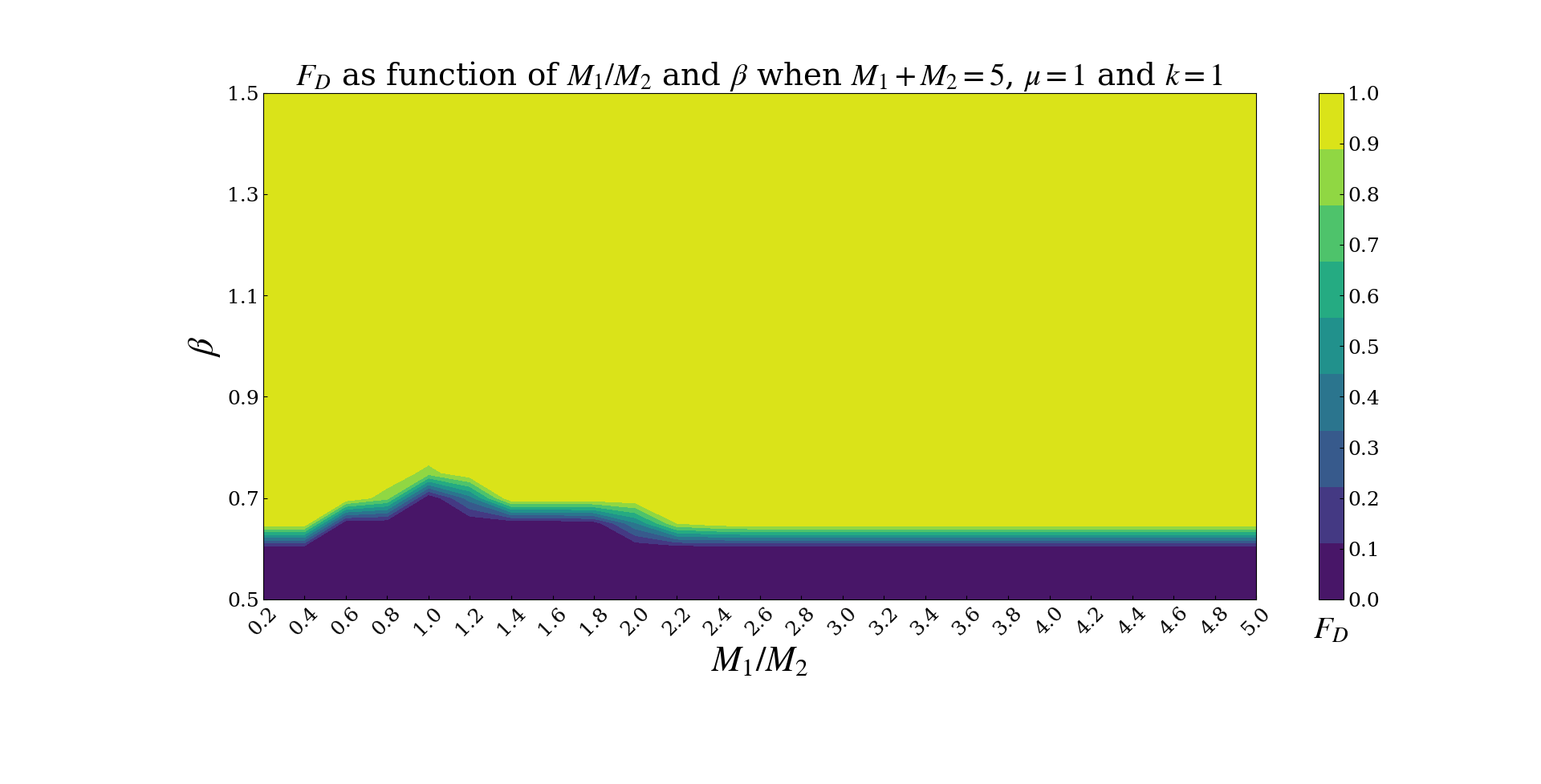}
		\caption{\(F_D\), \(M_1+M_2=5.0\)}
		\label{fig:beta_05a}
	\end{subfigure}
	\hfill
	\begin{subfigure}[b]{0.48\textwidth}
		\centering
		\includegraphics[width=\textwidth]{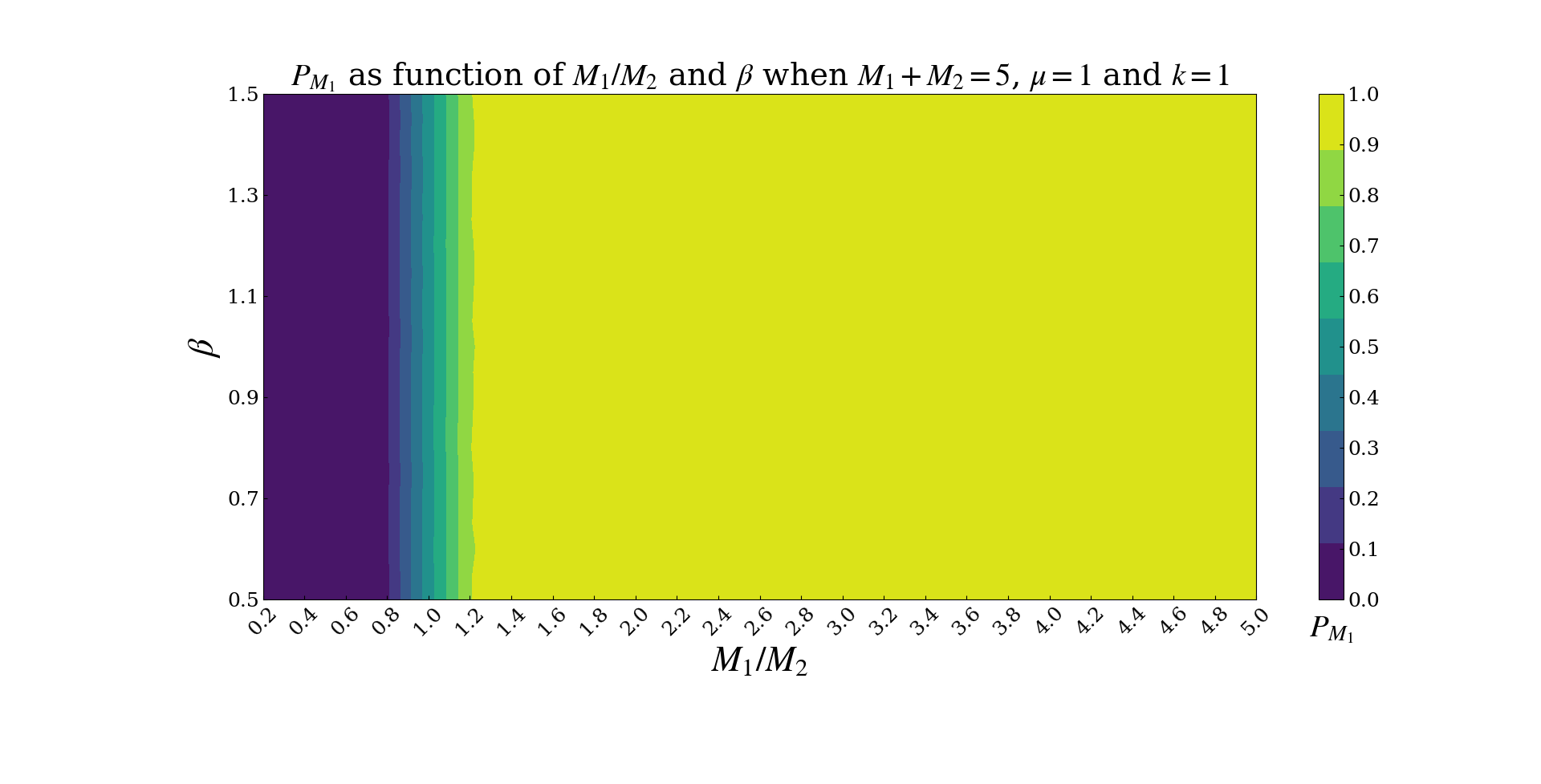}
		\caption{\(P_{M_1}\), \(M_1+M_2=5.0\)}
		\label{fig:beta_05b}
	\end{subfigure}
	
	\caption{Effect of utility multiplier \(\beta\) on involution and spatial distribution for three total resource levels. Each pair of panels corresponds to a fixed total resource: (a,b) \(M_1+M_2=2.0\), (c,d) \(M_1+M_2=3.0\), (e,f) \(M_1+M_2=5.0\). Simulations with \(k=1.0\), \(\mu=1.0\), 2000 agents, averaged over 50 independent runs. The standard deviation of $F_D$ and $P_{M_1}$ across all simulation results is below 0.05.}\label{fig:beta}
\end{figure}

The parameter \(\beta\) scales the utility of high effort relative to low effort, with \(u_D = \beta e_D\) and \(e_D = 0.2\) fixed. We systematically varied \(\beta\) from \(0.5\) to \(1.5\) while keeping other parameters constant (\(k=1.0\), \(\mu=1.0\), 2000 agents). Figure~\ref{fig:beta} presents the results for three total resource levels (\(M_1+M_2 = 2.0, 3.0, 5.0\)), each showing both the fraction of high-effort agents \(F_D\) and the fraction of agents located in region \(M_1\), \(P_{M_1}\), as functions of the resource ratio \(M_1/M_2\) and \(\beta\).

Figure~\ref{fig:beta} reveals that \(\beta\) acts as a key tuning parameter for involution, with its effect strongly modulated by total resource availability. At low total resources (\(M_1+M_2 = 2.0\), panels a and b), \(\beta\) must exceed a threshold of approximately \(0.85\) for high-effort adoption to emerge at all. Below this threshold, \(F_D\) remains zero across all resource ratios, indicating that the payoff advantage of high effort is insufficient to overcome its cost, even under extreme resource asymmetry. Once \(\beta\) surpasses this threshold, \(F_D\) begins to rise, but only for sufficiently large \(M_1/M_2\) (or its reciprocal). For example, at \(\beta = 0.9\), \(F_D\) exceeds \(0.5\) only when \(M_1/M_2 > 3.0\) or \(< 0.33\); at \(\beta = 1.0\), the threshold shifts to approximately \(M_1/M_2 > 2.0\) or \(< 0.5\). At the highest \(\beta\) values (\(\geq 1.3\)), \(F_D\) approaches \(1.0\) for all but the most balanced ratios near unity. The spatial distribution \(P_{M_1}\) mirrors this behavior: at low \(\beta\), agents remain nearly evenly distributed (\(P_{M_1} \approx 0.5\)) regardless of resource ratio, because the low payoff of high effort discourages migration to the richer region. As \(\beta\) increases, agents concentrate more strongly in the resource-rich region, with \(P_{M_1}\) approaching \(1.0\) for \(M_1/M_2 > 1\) and \(0.0\) for \(M_1/M_2 < 1\), and the transition sharpens with increasing \(\beta\).

At moderate total resources (\(M_1+M_2 = 3.0\), panels c and d), the system exhibits a more abrupt transition. For \(\beta \leq 0.7\), \(F_D\) is zero across all ratios, again indicating that low \(\beta\) suppresses involution entirely. At \(\beta = 0.75\), \(F_D\) jumps to near-unity for extreme ratios (\(M_1/M_2 > 3.0\) or \(< 0.33\)), while remaining low near symmetry. As \(\beta\) increases further, the range of ratios supporting high \(F_D\) expands rapidly: at \(\beta = 0.8\), \(F_D > 0.9\) for all \(M_1/M_2 \geq 1.8\) or \(\leq 0.56\); at \(\beta = 0.85\), the region of low \(F_D\) shrinks to a narrow band around \(M_1/M_2 = 1\); and for \(\beta \geq 0.9\), \(F_D\) is uniformly \(1.0\) across the entire parameter space. This indicates a sharp threshold: once \(\beta\) exceeds a critical value (approximately \(0.85\) at this resource level), the entire population adopts the high-effort strategy regardless of resource distribution. The spatial distribution \(P_{M_1}\) shows a corresponding pattern: at low \(\beta\), agents are evenly spread; as \(\beta\) crosses the threshold, concentration in the richer region becomes extreme, with \(P_{M_1}\) approaching its extreme values rapidly, confirming that the decisions to adopt high effort and to migrate are tightly coupled.

When resources are abundant (\(M_1+M_2 = 5.0\), panels e and f), the influence of \(\beta\) is less dramatic but still discernible. At the lowest \(\beta\) values (\(0.5\)-\(0.6\)), \(F_D\) is already non-zero for moderately asymmetric ratios: for example, at \(\beta = 0.5\), \(F_D \approx 0.5\) at \(M_1/M_2 = 1.0\) and rises to near \(1.0\) for ratios above \(1.5\) or below \(0.67\). As \(\beta\) increases, the range of ratios supporting high \(F_D\) expands, and the transition becomes sharper. At \(\beta = 0.7\), \(F_D\) is already \(1.0\) for all ratios except those very close to unity. For \(\beta \geq 0.75\), \(F_D\) is uniformly \(1.0\) across the entire parameter space, indicating that abundance combined with even a modest \(\beta\) suffices to drive universal involution. The spatial distribution \(P_{M_1}\) shows that even at the lowest \(\beta\), agents concentrate strongly in the richer region when \(M_1/M_2\) deviates from unity, reflecting the strong pull of abundant resources regardless of the payoff multiplier. As \(\beta\) increases, this concentration becomes even more pronounced, with \(P_{M_1}\) approaching its extreme values more rapidly.

Taken together, these results demonstrate that \(\beta\) serves as a critical determinant of involution, with its effect exhibiting a clear threshold behavior that depends on total resource availability. Low \(\beta\) (below approximately \(0.7\)-\(0.8\)) suppresses high-effort adoption entirely unless resources are both abundant and highly asymmetric. As \(\beta\) increases, a sharp transition occurs beyond which the population switches abruptly to universal high effort, with the threshold occurring at lower \(\beta\) when resources are more abundant. The spatial concentration of agents in the richer region closely tracks this strategy shift, indicating that migration and strategy choice are mutually reinforcing. These findings underscore that policies aimed at reducing involution should consider not only the total resource pool and its spatial distribution, but also the effective payoff multiplier---which in real-world contexts could be influenced by factors such as tax rates, wage floors, or effort caps.

\subsection{Effect of effort ratio $e_2/e_1$}
\label{subsec:effort_ratio}

\begin{figure}[h]
	\centering
	\begin{subfigure}[b]{0.48\textwidth}
		\includegraphics[width=\textwidth]{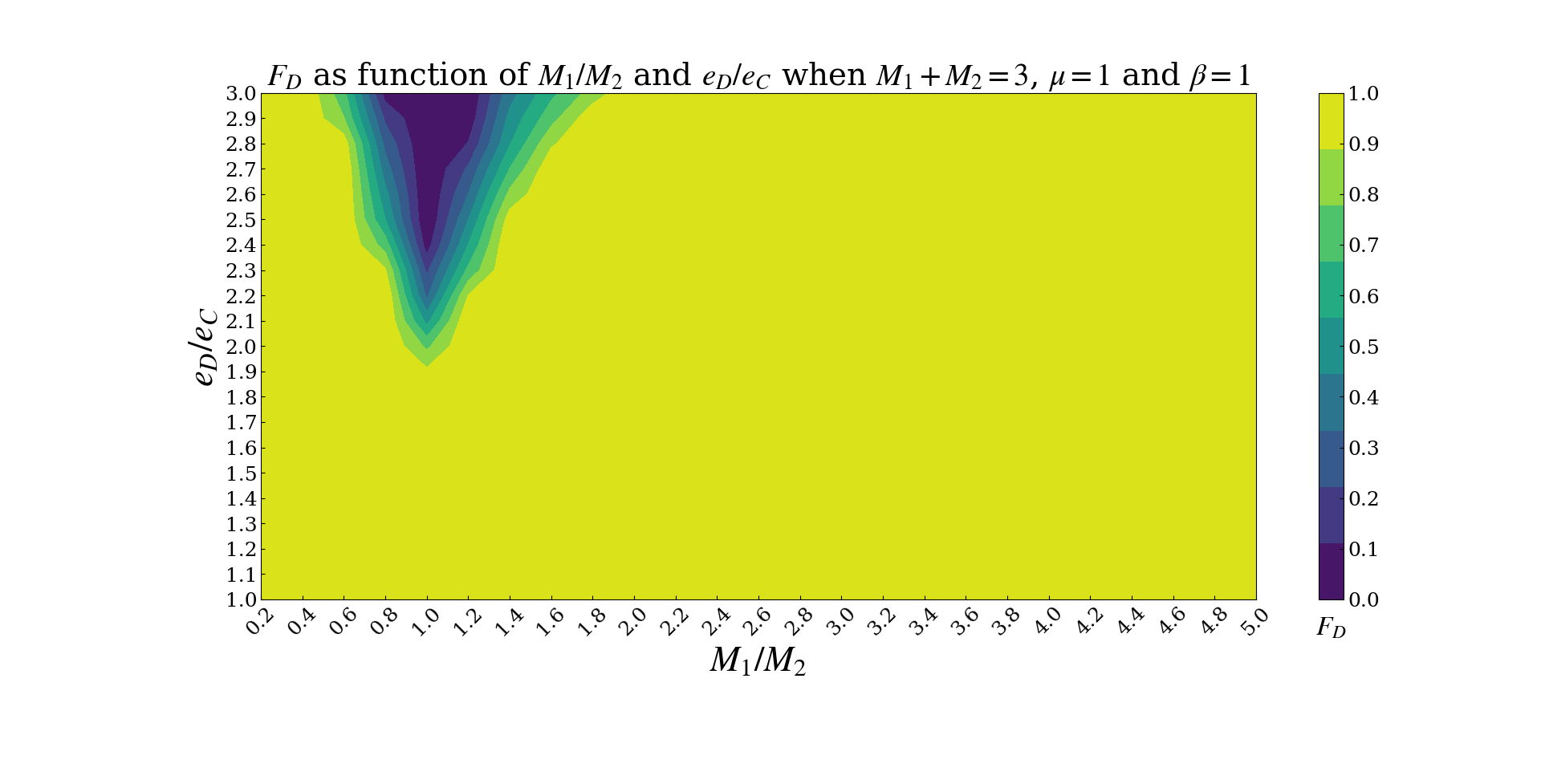}
		\caption{\(F_D\) vs.\ \(e_2/e_1\) for various \(M_1/M_2\)}
		\label{fig:e1e2a}
	\end{subfigure}
	\hfill
	\begin{subfigure}[b]{0.48\textwidth}
		\includegraphics[width=\textwidth]{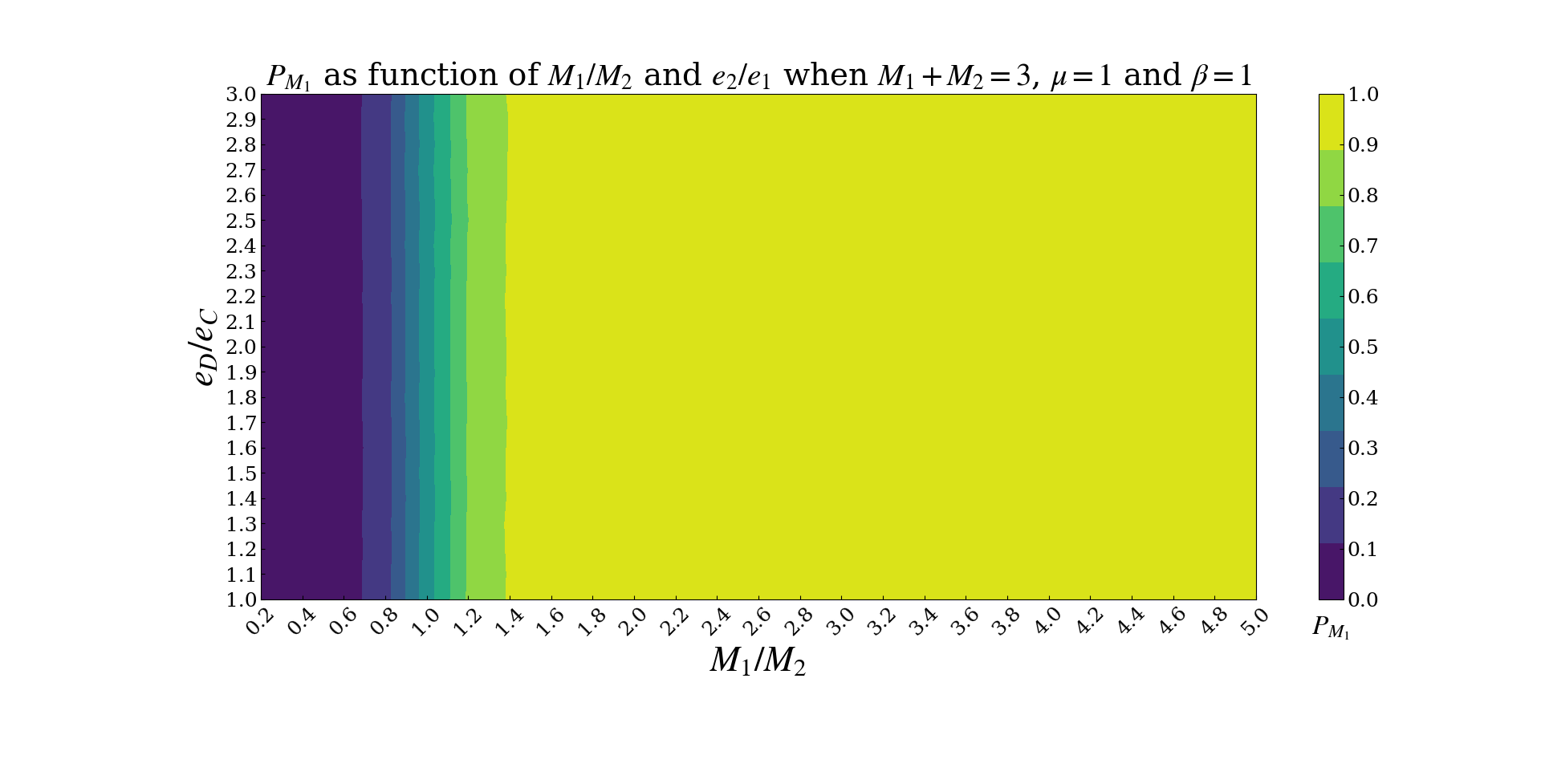}
		\caption{\(F_D\) vs.\ \(M_1/M_2\) for various \(e_D/e_C\)}
		\label{fig:e1e2b}
	\end{subfigure}
	\caption{Effect of the effort ratio \(e_2/e_1\) on involution. Simulations with \(M_1+M_2 = 3.0\), \(\mu = 1.0\), \(k = 1.0\), and 2000 agents, averaged over 50 independent runs. The standard deviation of $F_D$ and $P_{M_1}$ across all simulation results is below 0.05.}\label{fig:e1e2}
\end{figure}

We also varied the effort levels while keeping utilities proportional to efforts (i.e., \(u_i = e_i\)). Figure~\ref{fig:e1e2} presents the results for different effort ratios \(e_2/e_1\) (where \(e_1 = e_C\), \(e_2 = e_D\)) under fixed total resources \(M_1+M_2 = 3.0\), migration rate \(\mu = 1.0\), and selection intensity \(k = 1.0\). Panel (a) shows the fraction of high-effort agents \(F_D\) as a function of the effort ratio for several representative resource ratios \(M_1/M_2\), while panel (b) displays \(F_D\) as a function of \(M_1/M_2\) for various effort ratios.

Figure~\ref{fig:e1e2} reveals that the effort ratio \(e_2/e_1\) critically modulates the conditions under which involution emerges. When the effort ratio is close to unity (\(e_2/e_1 \approx 1.0\)), high effort is only slightly more costly than low effort, and \(F_D\) remains uniformly high (near \(1.0\)) across all resource ratios \(M_1/M_2\). As the effort ratio increases, making high effort progressively more expensive, a distinct trough appears around the symmetric point \(M_1/M_2 = 1.0\). For instance, at \(e_2/e_1 = 2.0\), \(F_D\) drops to approximately \(0.75\) at \(M_1/M_2 = 1.0\), while still exceeding \(0.99\) at \(M_1/M_2 = 0.8\) and \(1.2\). As the effort ratio rises further, the central trough deepens and widens: at \(e_2/e_1 = 2.5\), \(F_D\) falls below \(0.01\) at the symmetric point, but remains above \(0.94\) at \(M_1/M_2 = 1.4\); at \(e_2/e_1 = 3.0\), \(F_D\) is essentially zero for \(0.8 \lesssim M_1/M_2 \lesssim 1.2\), while approaching unity for ratios below \(0.6\) or above \(1.6\). The curves exhibit clear symmetry about \(M_1/M_2 = 1\) (on a logarithmic scale), confirming that the system responds identically to swapping the two regions.

The spatial distribution \(P_{M_1}\) (data not shown) mirrors these trends. For low effort ratios, agents concentrate strongly in the richer region whenever \(M_1/M_2 \neq 1\). As the effort ratio increases, the degree of concentration near the symmetric point diminishes, with \(P_{M_1}\) approaching \(0.5\) for \(M_1/M_2 \approx 1\), indicating that the high cost of effort discourages migration when resource gains are marginal. However, for extreme resource ratios, \(P_{M_1}\) remains close to \(0\) or \(1\), reflecting the persistent pull of abundant resources.

These results demonstrate that the effort ratio sets a threshold for involution: the greater the relative cost of high effort, the larger the resource asymmetry required to sustain it. In policy terms, measures that increase the effective cost of excessive effort (e.g., progressive taxation on overtime, strict working hour limits) can curb involution by raising the bar for when competition becomes worthwhile, even without altering the underlying resource distribution.

\subsection{Influence of resource distribution patterns}
\label{subsec:resource_pattern}

\begin{figure}[h]
	\centering
	\begin{subfigure}[b]{0.8\textwidth}
		\includegraphics[width=\textwidth]{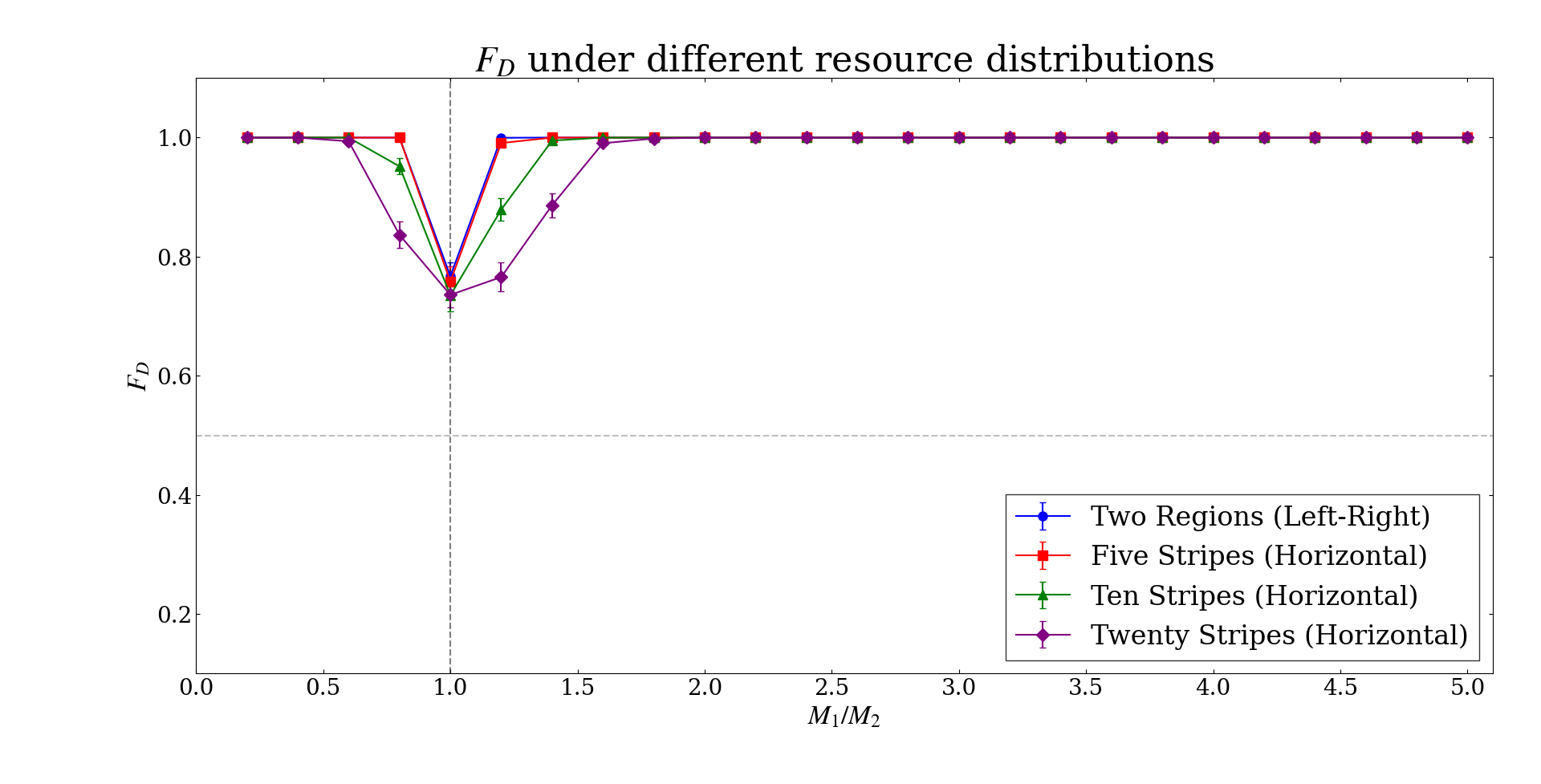}
		\caption{Fraction of high‑effort agents $F_D$ as a function of $M_1/M_2$ for different stripe patterns.}\label{fig:stripes_a}
	\end{subfigure}
	\hfill
	\begin{subfigure}[b]{0.8\textwidth}
		\includegraphics[width=\textwidth]{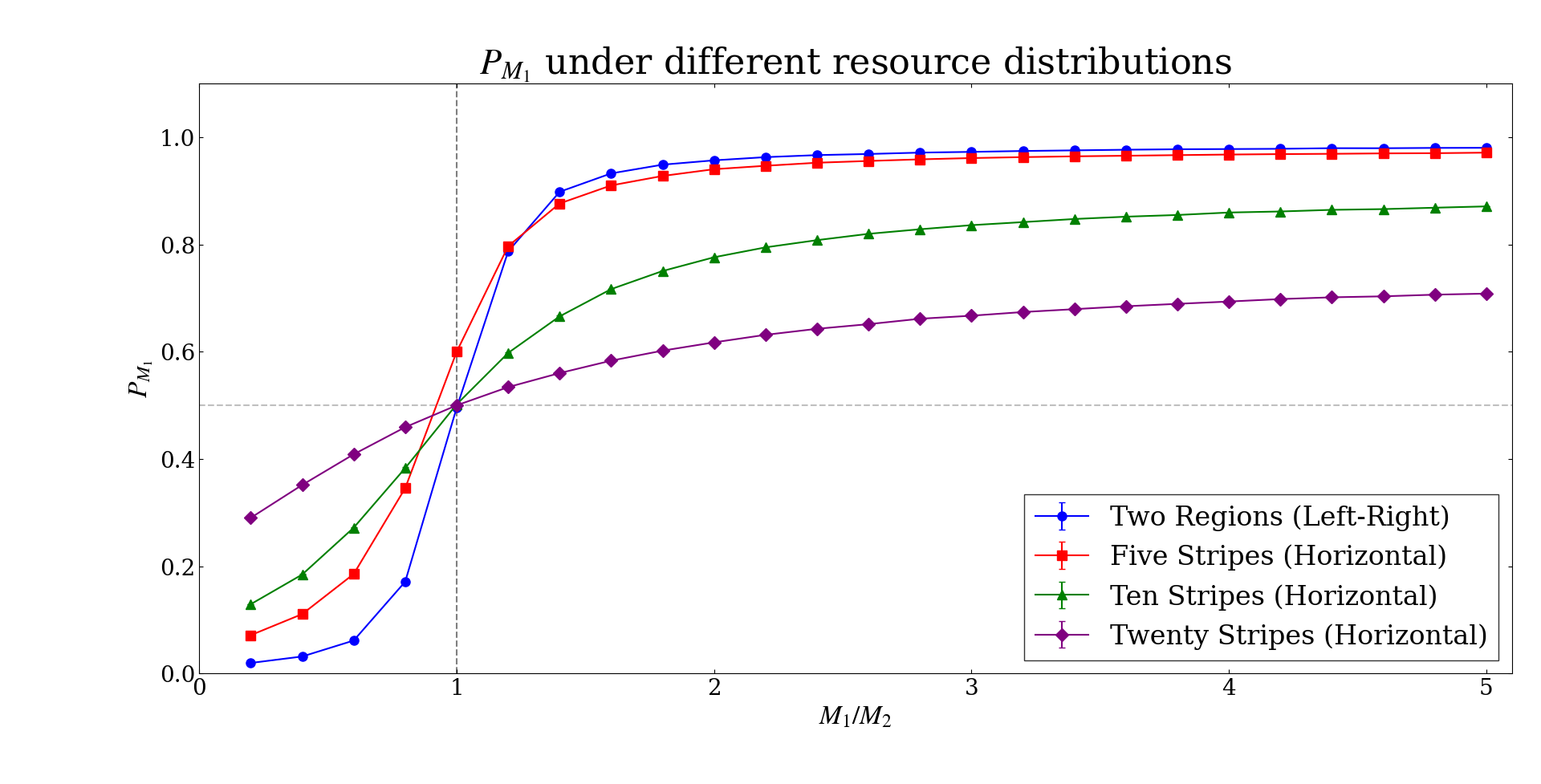}
		\caption{Fraction of agents located in the $M_1$ region $P_{M_1}$ as a function of $M_1/M_2$ for different stripe patterns.}\label{fig:stripes_b}
	\end{subfigure}
	\caption{Effect of resource mixing on involution and spatial distribution. Simulations with total resources $M_1+M_2=3.0$, migration rate $\mu=1.0$, selection intensity $k=1.0$, and 2000 agents. Each curve is averaged over 50 independent runs. Error bars are shown in the figures.}\label{fig:stripes}
\end{figure}

Resource distribution is not always a simple step function. 
To explore the effect of spatial mixing, we simulated configurations with alternating stripes of two resource levels. 
The number of stripes (two, five, ten, twenty) represents the degree of mixing: more stripes correspond to finer-grained alternation and thus higher spatial heterogeneity at the local scale. 
Figure~\ref{fig:stripes} presents the results.

Figure~\ref{fig:stripes_a} shows that increasing the degree of mixing (more stripes) systematically reduces the prevalence of the high‑effort strategy for a given resource ratio, except in the immediate vicinity of $M_1/M_2=1$ where $F_D$ remains largely unchanged. 
For example, at $M_1/M_2=0.8$, $F_D$ drops from nearly 1.0 in the two‑region case to about 0.87 for twenty stripes; a similar decline is observed at $M_1/M_2=1.2$. This indicates that spatial fragmentation mitigates involution by providing agents with more balanced local environments, thereby weakening the incentive to escalate effort.

Figure~\ref{fig:stripes_b} examines how mixing affects agent distribution. 
In the two‑region setup, agents strongly concentrate in the richer region: when $M_1/M_2<1$, $P_{M_1}$ is very low (most agents leave the poorer $M_1$ region); when $M_1/M_2>1$, $P_{M_1}$ approaches 1. 
As stripes become finer, this polarisation diminishes: $P_{M_1}$ moves closer to 0.5 for all resource ratios, indicating a more uniform spatial distribution. 
Finer mixing thus reduces the tendency of agents to flock to the resource‑rich area, which in turn lowers local competitive pressure and contributes to the suppression of $F_D$ seen in panel (a).

Together, these results demonstrate that even without altering the global resource totals, the way resources are arranged spatially can significantly influence both the intensity of involution and the distribution of agents. 
Policies that promote mixing of different resource levels (e.g., through urban planning or economic diversification) may therefore help alleviate excessive competition.

\subsection{Impact of agent population size}
\label{subsec:agent_number}
Before varying population density, we verified that the final outcomes are insensitive to how agents are initially placed on the lattice. In addition to the default random initial positions, we tested two extreme configurations: (i) \textit{clustered} (all agents initially in a single contiguous block covering 20\% of the lattice), and (ii) \textit{stratified} (agents evenly split between the two resource regions, each at density 0.2). Using the same parameters as in Fig.~2 (\(M_1+M_2=3.0\), \(M_1/M_2=2.0\), \(\mu=1.0\), \(k=1.0\)), all three conditions yielded nearly identical final fractions of high-effort agents (\(F_D\)) and spatial distributions (\(P_{M_1}\)), with differences well within statistical fluctuations (e.g., \(P_{M_1} \approx 0.957 \pm 0.004\) in all cases). This robustness arises because migration and strategy updating rapidly erase initial spatial correlations. Hence, the results presented below are not biased by the particular initial placement.

\begin{figure}[h]
	\centering
	\begin{subfigure}[b]{0.8\textwidth}
		\includegraphics[width=\textwidth]{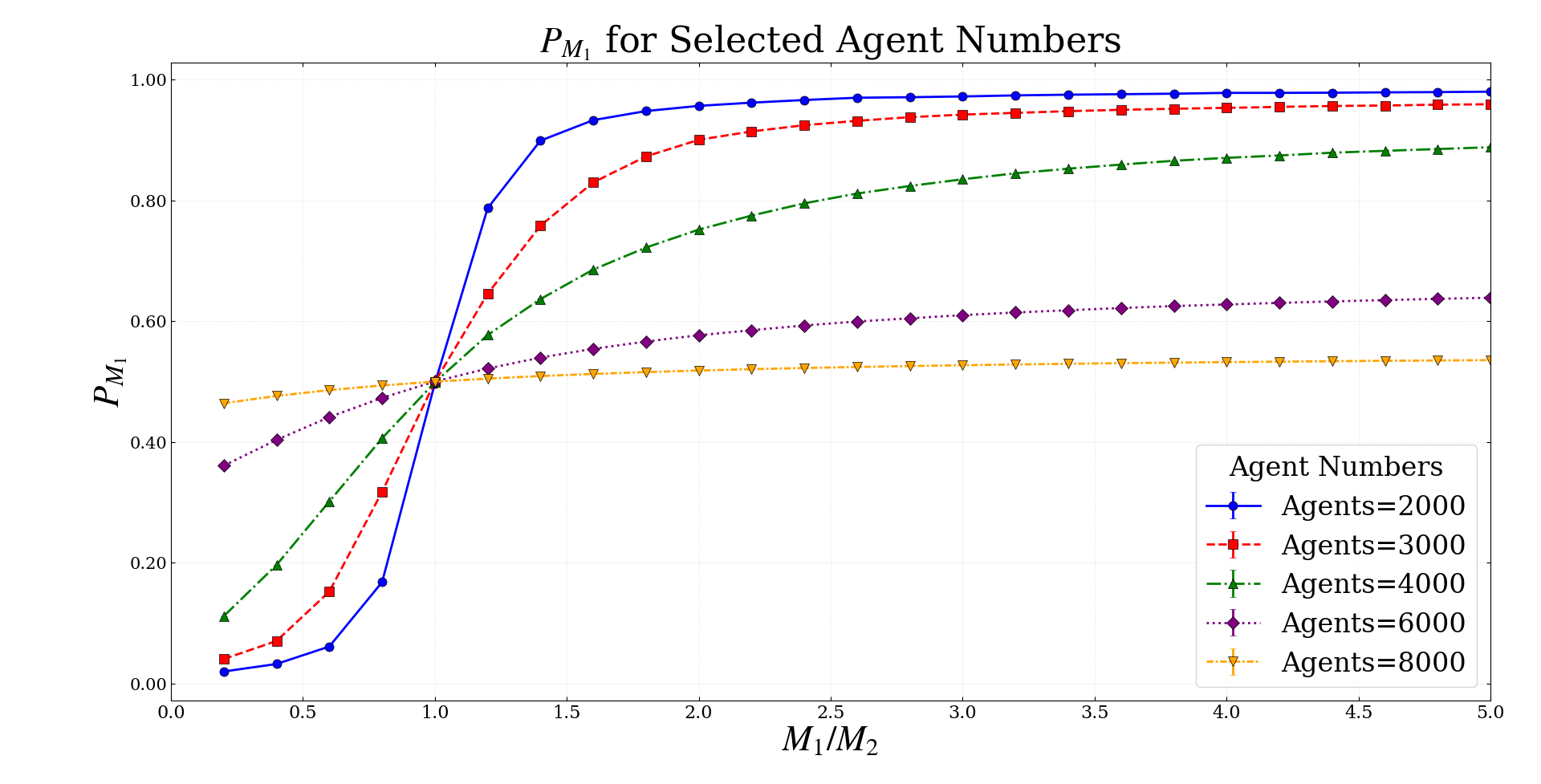}
		\caption{\(P_{M_1}\) vs.\ \(M_2/M_1\) for different agent numbers}\label{fig:agents_a}
	\end{subfigure}
	\hfill
	\begin{subfigure}[b]{0.8\textwidth}
		\includegraphics[width=\textwidth]{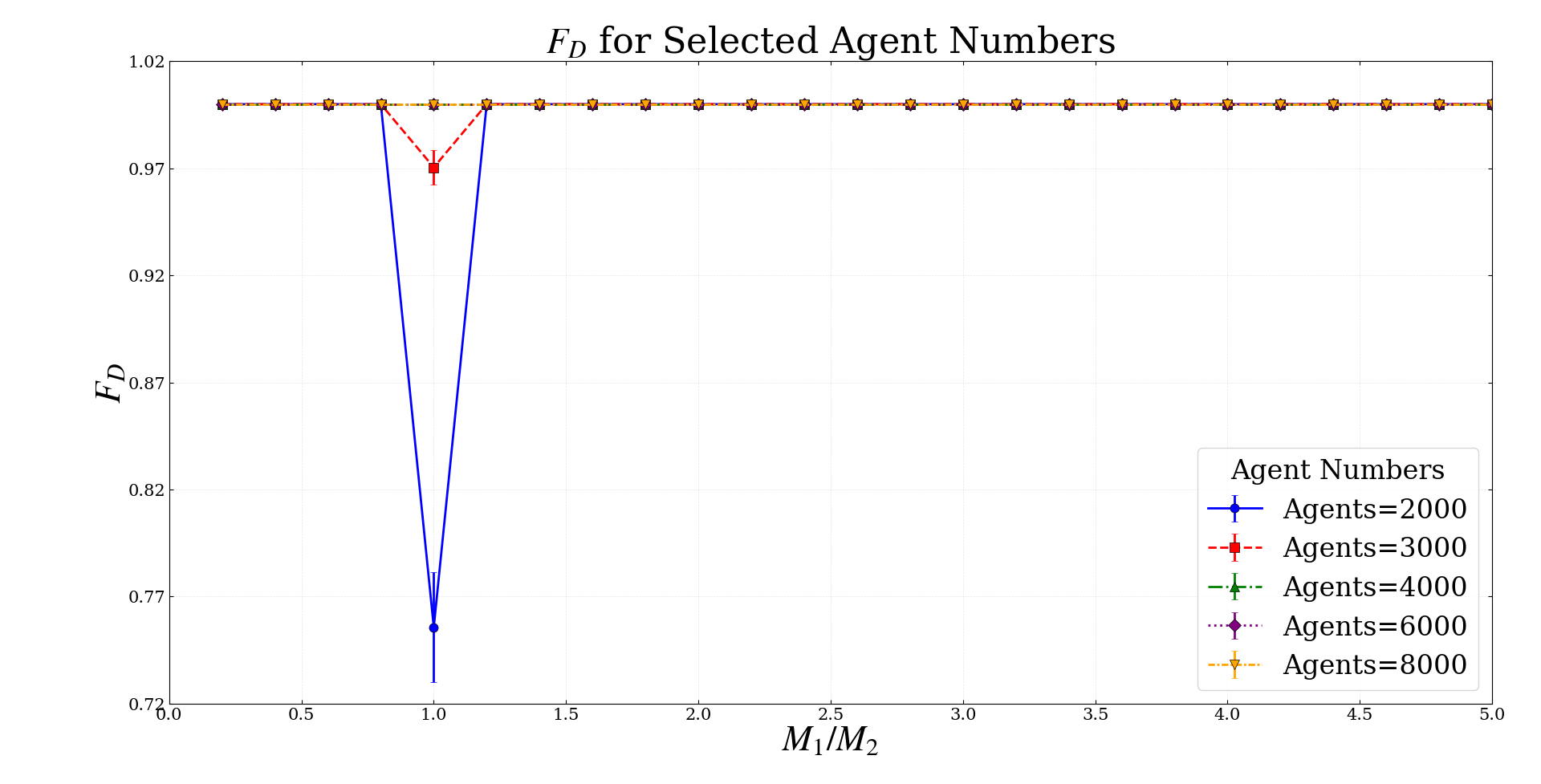}
		\caption{\(F_D\) vs.\ agent number for fixed \(M_2/M_1\) ratios}\label{fig:agents_b}
	\end{subfigure}
	\caption{Effect of population size on spatial distribution and involution. Simulations with total resources \(M_1+M_2=3.0\), migration rate \(\mu=1.0\), selection intensity \(k=1.0\), and 50 independent runs per condition. Error bars are shown in the figures.}\label{fig:agents}
\end{figure}

The number of agents (i.e., population density) modulates the intensity of local competition and spatial sorting. Figure~\ref{fig:agents} examines the effect of varying the total number of agents from 2000 to 8000 (corresponding to densities \(\rho = 0.2\) to \(0.8\)) while keeping the lattice size fixed (\(100\times100\)). Panel (a) shows the fraction of agents located in the \(M_1\) region, \(P_{M_1}\), as a function of the resource ratio \(M_2/M_1\) for different agent numbers. Panel (b) displays the fraction of high-effort agents, \(F_D\), as a function of agent number for several representative values of \(M_2/M_1\), highlighting how population density influences the propensity to adopt the competitive strategy.

Figure~\ref{fig:agents_a} reveals that spatial concentration in the richer region is strongly affected by both resource asymmetry and population density. 
When \(M_2/M_1 < 1\) (i.e., \(M_1\) is the poorer region), \(P_{M_1}\) is very low for small populations, indicating that most agents abandon the poorer region. 
As density increases, however, \(P_{M_1}\) rises markedly, approaching 0.5 for the highest densities even at extreme ratios. 
For example, at \(M_2/M_1=0.2\) (i.e., \(M_1/M_2=5\)), \(P_{M_1}\) increases from about 0.02 for 2000 agents to nearly 0.46 for 8000 agents. 
This trend demonstrates that crowding in the resource-rich region forces some agents to remain in or move to the poorer area, thereby reducing spatial inequality. 
The same pattern holds symmetrically when \(M_2/M_1 > 1\) (with \(M_1\) being the richer region), where \(P_{M_1}\) decreases toward 0.5 as density grows.

Figure~\ref{fig:agents_b} illustrates how population density affects the adoption of the high-effort strategy. 
For resource ratios sufficiently far from unity (e.g., \(M_2/M_1 = 0.5\) or \(2.0\)), \(F_D\) is uniformly 1.0 regardless of agent number, indicating that extreme resource asymmetry alone suffices to drive the entire population toward involution. 
In contrast, near the symmetric point \(M_2/M_1 = 1.0\), \(F_D\) exhibits a clear dependence on density: it increases from approximately 0.75 for 2000 agents to 0.99 for 3000 agents and reaches 1.0 for larger populations. 
Thus, higher population density intensifies competition even when resources are evenly distributed, pushing the system toward universal high effort.

Taken together, these results show that population density has contrasting effects on spatial distribution and strategic choice. On one hand, higher density promotes a more balanced spatial distribution by saturating attractive regions and forcing agents to occupy less favorable areas. 
On the other hand, it exacerbates involution by increasing local competitive pressure, particularly in situations where resources are nearly equal. 
The net outcome is that while agents become more evenly spread, they all tend to adopt the high-effort strategy, leading to widespread involution irrespective of spatial equity.

\subsection{Synthesis and policy implications}
\label{subsec:synthesis}

Collectively, our results reveal a nuanced picture: migration generally suppresses global involution by mixing strategies, but it also concentrates agents in resource-rich areas, creating local hot spots of competition. 
The net effect depends on the scale of observation. 
Policies aiming to reduce involution should consider both mobility enhancement and resource redistribution. In particular:
\begin{itemize}
	\item Enhancing mobility (e.g., through job platforms, relocation subsidies) can help agents escape saturated markets, but without addressing resource disparities it merely shifts competition.
	\item Reducing resource inequality between regions prevents the formation of extreme concentration and the associated local involution.
	\item Fragmentation of resource patches (e.g., mixing high‑ and low-resource zones) can lower involution even without changing total resources.
	\item Policy interventions should be tailored to the resource context: mobility works best when resources are adequate, but may have limited impact under extreme scarcity.
\end{itemize}

\section{Theoretical Analysis}\label{sec:theory}
Mean-field theory is a standard tool from statistical physics and has been applied in evolutionary game theory for nearly two decades, with early applications to spatial games dating back to the mid-2000s \cite{szabo2005phase, Vukov20026}.
To gain deeper insight into the simulation results, we develop a mean-field approximation of the model. This approach neglects spatial correlations and local fluctuations, assuming that agents are uniformly distributed within each region and that interactions are well-mixed. Despite its simplifications, the mean-field theory captures the essential mechanisms driving migration and strategy evolution, and provides analytical predictions for the asymptotic behavior of the system.

\subsection{Mean-Field Payoffs}

Let $\rho = N_{\text{agents}}/N^2$ denote the global density of agents. Define $F_D$ as the fraction of agents adopting the high-effort strategy D, and $P_{M_1}$ as the fraction of agents located in region $M_1$ (the complementary fraction $1-P_{M_1}$ resides in region $M_2$). The average utility in the whole population is
\begin{equation}
	\bar{u} = F_D u_D + (1-F_D) u_C,
\end{equation}
where $u_C = e_C$ and $u_D = \beta e_D$, with $\beta$ being the utility multiplier for the high-effort strategy.

Consider an agent of type $i \in \{C,D\}$ located in region $r \in \{1,2\}$ with resource level $M_r$. Under the mean-field approximation, the expected payoff per competition (i.e., the average over the five lattice sites the agent participates in) is given by
\begin{equation}
	\mathbb{E}[\pi_i^r] = M_r \frac{u_i}{u_i + 4\rho \bar{u}} - e_i.
	\label{eq:indiv_payoff_mf}
\end{equation}
The derivation is as follows: at each of the five sites, the agent competes with the occupants of the four neighboring sites. The expected total utility competing at that site is $u_i$ (the focal agent's own utility) plus $4\rho\bar{u}$ (the expected utility from the four neighbors, each occupied with probability $\rho$ and having average utility $\bar{u}$). 
Hence the agent's share of the resource $M_r$ is $M_r \cdot u_i/(u_i+4\rho\bar{u})$. 
Summing over five sites and dividing by five gives the average payoff per competition. 
The cost term $-e_i$ reflects that the agent expends its full effort $e_i$ in each competition, resulting in a total cost of $5e_i$ over five competitions, which averages to $e_i$ per competition.

The average payoff in region $r$ is obtained by averaging over the two strategies:
\begin{equation}
	\mathbb{E}[\pi_r] = F_D \mathbb{E}[\pi_D^r] + (1-F_D) \mathbb{E}[\pi_C^r].
\end{equation}
Substituting Eq.~\eqref{eq:indiv_payoff_mf} yields
\begin{equation}
	\mathbb{E}[\pi_r] = M_r H(\bar{u}) - \big[F_D e_D + (1-F_D) e_C\big],
	\label{eq:region_payoff_mf}
\end{equation}
where we have introduced the auxiliary function
\begin{equation}
	H(\bar{u}) = F_D \frac{u_D}{u_D + 4\rho \bar{u}} + (1-F_D) \frac{u_C}{u_C + 4\rho \bar{u}}.
\end{equation}
Since $u_D > u_C > 0$, we have $H(\bar{u}) > 0$ for all $\bar{u} > 0$.

\subsection{Migration Dynamics}

The migration process described in Sec.~\ref{sec:model} implies that agents compare payoffs and tend to move toward regions offering higher returns. 
Under the mean-field approximation, the evolution of the spatial distribution $P_{M_1}$ can be described by the following differential equation \cite{sigmund_evolutionary_1998, hauert_game_2005, nowak_evolutionary_2006}:
\begin{equation}
	\frac{dP_{M_1}}{dt} = \mu P_{M_1}(1-P_{M_1}) \tanh\!\left[ \frac{k}{2}\big( \mathbb{E}[\pi_{M_1}] - \mathbb{E}[\pi_{M_2}] \big) \right],
	\label{eq:mig_dyn_mf}
\end{equation}
where $\mu$ is the migration rate and $k$ the selection intensity. 
This form arises from the pairwise comparison rule: the probability that an agent in the poorer region moves to the richer region is proportional to $\tanh$ of the payoff difference, and the factor $P_{M_1}(1-P_{M_1})$ captures the fraction of agents that are potentially mobile (those in the less attractive region) and the availability of empty sites in the more attractive region.

Using Eq.~\eqref{eq:region_payoff_mf}, the payoff difference simplifies to
\begin{equation}
	\mathbb{E}[\pi_{M_1}] - \mathbb{E}[\pi_{M_2}] = (M_1 - M_2) H(\bar{u}),
	\label{eq:payoff_diff_reg}
\end{equation}
because the cost terms cancel. Substituting into Eq.~\eqref{eq:mig_dyn_mf} gives
\begin{equation}
	\frac{dP_{M_1}}{dt} = \mu P_{M_1}(1-P_{M_1}) \tanh\!\left[ \frac{k}{2} (M_1 - M_2) H(\bar{u}) \right].
	\label{eq:mig_dyn_final}
\end{equation}

Since $H(\bar{u}) > 0$, the sign of the argument inside the hyperbolic tangent is determined solely by $M_1 - M_2$. Consequently, we obtain the following result:

\begin{proposition}[Migration equilibrium]
	Under the mean-field dynamics \eqref{eq:mig_dyn_final}, the system exhibits:
	\begin{itemize}
		\item If $M_1 > M_2$, then $dP_{M_1}/dt > 0$ for all $0 < P_{M_1} < 1$, implying that $P_{M_1}$ increases monotonically and converges to $1$ (all agents concentrate in region $M_1$).
		\item If $M_1 < M_2$, then $dP_{M_1}/dt < 0$ for all $0 < P_{M_1} < 1$, and $P_{M_1}$ decreases to $0$ (all agents concentrate in region $M_2$).
		\item If $M_1 = M_2$, then $dP_{M_1}/dt = 0$ for any $P_{M_1} \in [0,1]$; the system is neutrally stable and any initial distribution persists.
	\end{itemize}
	Thus, any asymmetry in regional resources inevitably drives the population toward complete agglomeration in the more resource-abundant region.
\end{proposition}

\subsection{Strategy Dynamics}

The evolution of the fraction $F_D$ of high-effort agents is governed by a replicator-type equation derived from the pairwise comparison rule \cite{sigmund_evolutionary_1998, hauert_game_2005, nowak_evolutionary_2006}:
\begin{equation}
	\frac{dF_D}{dt} = \alpha F_D (1-F_D) \tanh\!\left[ \frac{k}{2}\big( \mathbb{E}[\pi_D] - \mathbb{E}[\pi_C] \big) \right],
	\label{eq:strat_dyn_mf}
\end{equation}
where $\alpha$ is the strategy update rate (set to $1$ in simulations). The average payoff for a strategy $i$, $\mathbb{E}[\pi_i]$, is obtained by averaging over regions:
\begin{equation}
	\mathbb{E}[\pi_i] = P_{M_1} \mathbb{E}[\pi_i^{M_1}] + (1-P_{M_1}) \mathbb{E}[\pi_i^{M_2}].
\end{equation}
Using Eq.~\eqref{eq:indiv_payoff_mf} and introducing the region-averaged resource $\bar{M} = P_{M_1}M_1 + (1-P_{M_1})M_2$, we obtain
\begin{equation}
	\mathbb{E}[\pi_i] = \bar{M} \frac{u_i}{u_i + 4\rho \bar{u}} - e_i.
	\label{eq:strat_payoff_mf}
\end{equation}

Define the payoff difference $\Delta \pi = \mathbb{E}[\pi_D] - \mathbb{E}[\pi_C]$. From Eq.~\eqref{eq:strat_payoff_mf},
\begin{equation}
	\Delta \pi = \bar{M} \Phi(\bar{u}) - \Delta e,
	\label{eq:payoff_diff_strat}
\end{equation}
where $\Delta e = e_D - e_C > 0$ and
\begin{equation}
	\Phi(\bar{u}) = \frac{u_D}{u_D + 4\rho \bar{u}} - \frac{u_C}{u_C + 4\rho \bar{u}}
	= \frac{4\rho (u_D - u_C) \bar{u}}{(u_D + 4\rho \bar{u})(u_C + 4\rho \bar{u})}.
\end{equation}
Note that $\Phi(0)=0$, $\Phi(\bar{u})>0$ for $\bar{u}>0$, and $\Phi$ is a unimodal function achieving its maximum at $\bar{u}_{\text{max}} = \sqrt{u_C u_D}/(4\rho)$.

The fixed points of Eq.~\eqref{eq:strat_dyn_mf} are $F_D = 0$, $F_D = 1$, and any $F_D$ satisfying $\Delta \pi = 0$. The stability of these fixed points is determined by the sign of $\Delta \pi$ in their vicinity. 
Specifically, $F_D = 0$ is stable if $\Delta \pi < 0$ for $F_D$ near $0$, and $F_D = 1$ is stable if $\Delta \pi > 0$ for $F_D$ near $1$. An interior fixed point exists only when $\Delta \pi = 0$ has a solution $F_D^* \in (0,1)$; its stability depends on the derivative of $\Delta \pi$ with respect to $F_D$.

For the parameter values used in our simulations ($e_C = 0.1$, $e_D = 0.2$, $\beta = 1$, $\rho = 0.2$), we have $u_C = 0.1$, $u_D = 0.2$, and $4\rho\bar{u} = 0.8\bar{u}$. The maximum of $\Phi$ occurs at $\bar{u}_{\text{max}} = \sqrt{0.1\times 0.2}/0.8 \approx 0.1768$, with $\Phi_{\text{max}} \approx 0.1710$. At the boundaries,
\begin{align*}
	F_D = 0 &\Rightarrow \bar{u} = u_C = 0.1, \quad \Phi(0.1) \approx 0.1587,\\
	F_D = 1 &\Rightarrow \bar{u} = u_D = 0.2, \quad \Phi(0.2) \approx 0.1710.
\end{align*}
Thus $\Phi_{\text{min}} = \min\{\Phi(0.1),\Phi(0.2)\} = 0.1587$.

The interior equilibrium condition $\Delta \pi = 0$ reduces to
\begin{equation}
	\bar{M} \Phi(\bar{u}) = \Delta e = 0.1.
\end{equation}
Consequently, an interior equilibrium exists only when the average resource $\bar{M}$ lies in the interval
\begin{equation}
	\frac{0.1}{\Phi_{\text{max}}} \le \bar{M} \le \frac{0.1}{\Phi_{\text{min}}}
	\;\Longleftrightarrow\;
	0.585 \approx \frac{0.1}{0.1710} \le \bar{M} \le \frac{0.1}{0.1587} \approx 0.630.
	\label{eq:coexistence_interval}
\end{equation}

\begin{proposition}[Strategy equilibria]
	For the given parameters:
	\begin{itemize}
		\item If $\bar{M} < 0.585$, then $\Delta \pi < 0$ for all $F_D \in [0,1]$; hence $F_D = 0$ is globally stable (all agents adopt low effort).
		\item If $\bar{M} > 0.630$, then $\Delta \pi > 0$ for all $F_D \in [0,1]$; hence $F_D = 1$ is globally stable (all agents adopt high effort).
		\item If $0.585 < \bar{M} < 0.630$, there exist two interior fixed points: one unstable and one stable. The stable interior fixed point corresponds to a mixed-strategy equilibrium where both strategies coexist. However, due to the narrowness of this interval, such coexistence is rare and requires precise tuning of parameters.
	\end{itemize}
\end{proposition}

\subsection{Coupling of Migration and Strategy}

The two dynamical subsystems are coupled through the average resource $\bar{M} = P_{M_1}M_1 + (1-P_{M_1})M_2$, which depends on the spatial distribution, and the average utility $\bar{u}$, which depends on the strategy composition. 
Proposition 1 shows that migration drives $P_{M_1}$ to either $0$ or $1$ whenever $M_1 \neq M_2$. 
Consequently, $\bar{M}$ converges to the resource level of the richer region. 
Unless the two resource levels are very close (so that $\bar{M}$ falls inside the narrow coexistence interval), the strategy dynamics will then drive $F_D$ to the corresponding boundary equilibrium. 
This explains why the simulation results predominantly exhibit homogeneous strategy adoption (all $C$ or all $D$) when resource asymmetry is present. 
Only when resources are nearly equal and the migration-driven concentration is weak can mixed strategies persist.

These mean-field predictions are in excellent qualitative agreement with the simulation outcomes presented in Sec.~\ref{sec:results}, validating the theoretical framework and providing a mechanistic understanding of the observed involution dynamics.

\section{Discussion and Conclusion}
\label{sec:conclusion}

\subsection{Discussion}
This study has developed and analyzed an evolutionary game model of involution that incorporates agent migration and spatial resource heterogeneity. By integrating both strategy updating and mobility, we have extended existing involution models to better capture real-world competitive dynamics, such as those observed among food delivery riders and homogeneous stores. All simulation results are based on 50 independent runs, and error bars (standard deviation) are included in all figures (see Figs.~1,~2,~6,~7). Finite-size scaling tests confirmed that the chosen lattice size \(L=100\) is sufficient to avoid finite-size artifacts (see Model section).

Our theoretical analysis, based on mean-field approximations, revealed several key insights. First, migration dynamics are strongly driven by resource disparities: agents inevitably concentrate in regions with higher resource availability when $M_1 \neq M_2$. This finding aligns with empirical observations of labor and capital flows toward more prosperous areas. Second, strategy evolution exhibits threshold behavior: stable mixed-strategy equilibria (where both high- and low-effort strategies coexist) exist only within a narrow range of average resource levels $\bar{M}$. Outside this range, the population converges to either full cooperation (all low effort) or full involution (all high effort). 
The coupling between migration and strategy evolution means that resource heterogeneity often pushes the system toward extreme outcomes.
It is worth emphasizing that the mean-field analysis presented here serves as a baseline reference rather than a universal theory; its predictions are qualitatively valid on regular lattices but may not hold on all network topologies.

The simulation results corroborate these theoretical predictions. When resource differences are large, migration rapidly leads to regional concentration, and the subsequent strategy evolution typically results in homogeneous strategy adoption (either all $C$ or all $D$) within each region. Importantly, the presence or absence of migration---not its precise rate---is the critical factor: once migration is allowed, the exact value of the migration probability \(\mu\) has negligible impact on the final equilibrium, except near critical parameter values where stochastic effects and finite-size fluctuations become significant. The migration rate \(\mu\) primarily affects the speed of convergence rather than the final outcome.

Our work highlights the dual role of resource abundance. While increased resources can alleviate scarcity-driven competition, they may also intensify involution by raising the payoff differential between high- and low-effort strategies. This non‑monotonic relationship suggests that policies aimed solely at increasing resources may have unintended consequences. Instead, interventions that moderate the payoff advantage of excessive effort (e.g., through progressive taxation, effort caps, or minimum wage regulations) could be more effective in curbing involution.

Our findings also connect to two related concepts in the literature: the agglomeration dilemma and the migration dilemma. The agglomeration dilemma \cite{roca_emergence_2011} captures the tension between the benefits of spatial clustering (which can facilitate cooperation through network reciprocity) and the costs of overcrowding (which intensify local competition). In our model, migration drives agents to concentrate in resource-rich regions (Fig.~4), creating localized high-effort clusters — a pattern analogous to the agglomeration dilemma. The migration dilemma \cite{sadhukhan_cooperators_2021, sadhukhan_amplitude_2021} shows that cooperators may avoid migrating to defector-dominated demes due to fear of exploitation, yet migration is necessary for cooperation to spread across the population. In our involution game, agents face a cost‑benefit trade‑off when deciding whether to move toward successful individuals, and the mere presence of migration (regardless of its rate) dramatically alters evolutionary outcomes---echoing the insight from the migration dilemma literature that enabling mobility can fundamentally reshape population‑level dynamics. However, whereas the migration dilemma focuses on the barrier to cooperation posed by the risk of exploitation, our work highlights how migration interacts with spatial resource heterogeneity to modulate involution intensity. A systematic comparison of these three dilemmas---agglomeration, migration, and involution---represents a promising direction for future research.

Moreover, the model suggests that promoting mobility (e.g., through job‑matching platforms, relocation subsidies, or market information transparency) can help agents escape locally saturated markets. However, in the long run, unless resource disparities are addressed, mobility merely shifts the locus of competition rather than resolving it. Therefore, a combined approach that balances resource distribution and facilitates adaptive mobility may offer a more sustainable solution.

\subsection{Limitations and Future Work}

While our model captures essential features of involution and migration, several limitations warrant mention.

\textbf{First, mean-field idealization and network generality.} Our mean-field analysis assumes well-mixed conditions within each region and neglects spatial correlations and network effects. Although the mean-field predictions qualitatively match the simulation results on the regular lattice, this does not imply that the conclusions are universal across all network topologies. On networks with heterogeneous degree distributions (e.g., scale-free) or non-trivial clustering (e.g., small-world), spatial correlations and degree heterogeneity could significantly alter both strategy spread and migration patterns. Future work should therefore explore how network structure—such as small-world, scale-free, or adaptive networks—modulates the interplay between migration, resource heterogeneity, and involution. The mean-field analysis presented here should be viewed as a baseline reference rather than a universal theory.

\textbf{Second, simplified assumptions on resources and strategies.} The model assumes a fixed total population and a simple two-region resource distribution. Extending the model to multiple regions with dynamically changing resources (e.g., through economic growth or policy interventions) would provide a more nuanced understanding of long‑term dynamics. Additionally, we have considered only two discrete effort levels. A continuous strategy space, where agents can adjust effort incrementally, would allow for richer evolutionary trajectories and could reveal gradient dynamics of involution.

\textbf{Third, simplified migration behavior.} The current model assumes a fixed migration rate and does not include learning or adaptation of mobility. In reality, agents may adjust their propensity to move based on past experiences or social cues. Moreover, another important simplification is the absence of movement costs. In reality, migration---whether physical relocation or job switching---incurs time, energy, financial, or opportunity costs. Such costs can deter excessive mobility and may alter the spatial concentration patterns observed in our cost-free setting. As shown in \cite{10.1063/5.0100772}, even modest migration costs can qualitatively change the outcome of evolutionary games on networks. Extending our model to include a cost term (e.g., a fixed or payoff-dependent deduction when migrating) would likely reduce the tendency of agents to cluster in resource-rich areas, thereby moderating the local involution intensity. We therefore caution that our results are most directly applicable to contexts where migration costs are low (e.g., digital workers switching between online platforms, or street vendors moving short distances). Future work should systematically explore how migration costs reshape the phase diagram of involution.

\subsection{Conclusion}
In summary, we have presented an evolutionary game model that integrates strategy selection and migration in a spatially heterogeneous environment. 
The model demonstrates how resource distribution and mobility jointly shape involutionary outcomes. 
Theoretical analysis and simulations show that migration tends to amplify resource disparities, often driving populations toward extreme strategy profiles. 
The key policy-relevant insights are: (i) reducing regional resource inequality suppresses involution; (ii) increasing total resources can paradoxically exacerbate it; and (iii) enabling migration helps agents escape local saturation, but without addressing underlying resource disparities it merely relocates competition. 
These findings underscore the importance of considering spatial mobility in the study of involution and offer insights for designing policies that mitigate excessive competition while promoting efficient resource allocation.

By bridging the gap between evolutionary game theory and real-world socio-economic phenomena, this work contributes to a deeper understanding of involution dynamics and provides a foundation for further research on the complex interplay between competition, mobility, and resource heterogeneity.

\section*{Acknowledgements}
This work was supported by the Wuhan City Polytechnic Research Project (Grant No. 2025WHCPB02) and the National Natural Science Foundation of China (Grant No. 72231010).

\bibliographystyle{unsrtnat}
\bibliography{cas-refs}  

\begin{thebibliography}{41}
\providecommand{\natexlab}[1]{#1}
\providecommand{\url}[1]{\texttt{#1}}
\expandafter\ifx\csname urlstyle\endcsname\relax
  \providecommand{\doi}[1]{doi: #1}\else
  \providecommand{\doi}{doi: \begingroup \urlstyle{rm}\Url}\fi

\bibitem[Tao et~al.(2025)Tao, Lu, Shi, and Yuan]{TAO2025104531}
Yuan Tao, Yanping Lu, Wei Shi, and Guangzhe~Frank Yuan.
\newblock Multidimensional diagnostic features and intervention pathways for
  youth mental disorders in china’s involution society.
\newblock \emph{Asian Journal of Psychiatry}, 108:\penalty0 104531, 2025.
\newblock ISSN 1876-2018.
\newblock \doi{https://doi.org/10.1016/j.ajp.2025.104531}.
\newblock URL
  \url{https://www.sciencedirect.com/science/article/pii/S1876201825001741}.

\bibitem[Wang et~al.(2022)Wang, Huang, Pan, and He]{WANG2022112092}
Chaoqian Wang, Chaochao Huang, Qiuhui Pan, and Mingfeng He.
\newblock Modeling the social dilemma of involution on a square lattice.
\newblock \emph{Chaos, Solitons \& Fractals}, 158:\penalty0 112092, 2022.
\newblock ISSN 0960-0779.
\newblock \doi{https://doi.org/10.1016/j.chaos.2022.112092}.
\newblock URL
  \url{https://www.sciencedirect.com/science/article/pii/S0960077922003022}.

\bibitem[Xiong()]{xiong2025anti}
Yi~Xiong.
\newblock Understanding {China's} ``{Anti}-involution'' {Drive}.
\newblock Deutsche Bank Research Institute.
\newblock URL
  \url{https://www.dbresearch.com/PROD/RI-PROD/PDFVIEWER.calias?pdfViewerPdfUrl=PROD0000000000603307}.
\newblock Chief Economist, Ph.D.

\bibitem[Nowak and May(1992)]{nowak_evolutionary_1992}
Martin~A. Nowak and Robert~M. May.
\newblock Evolutionary games and spatial chaos.
\newblock \emph{Nature}, 359\penalty0 (6398):\penalty0 826--829, October 1992.
\newblock \doi{10.1038/359826A0}.
\newblock URL \url{https://www.nature.com/articles/359826a0}.

\bibitem[Szabó and Tőke(1997)]{szabo_evolutionary_1997}
György Szabó and Csaba Tőke.
\newblock Evolutionary prisoner's dilemma game on a square lattice.
\newblock \emph{Physical Review E}, 58\penalty0 (1):\penalty0 69--73, October
  1997.
\newblock \doi{10.1103/PHYSREVE.58.69}.
\newblock URL \url{https://doi.org/10.1103/PhysRevE.58.69}.

\bibitem[Szolnoki et~al.(2009{\natexlab{a}})Szolnoki, Perc, Szabó, and
  Stark]{szolnoki_impact_2009}
Attila Szolnoki, Matjaž Perc, György Szabó, and Hans-Ulrich Stark.
\newblock Impact of aging on the evolution of cooperation in the spatial
  prisoner's dilemma game.
\newblock \emph{Physical Review E}, 80\penalty0 (2):\penalty0 021901, August
  2009{\natexlab{a}}.
\newblock \doi{10.1103/PHYSREVE.80.021901}.
\newblock URL \url{https://doi.org/10.1103/PhysRevE.80.021901}.

\bibitem[Zhang et~al.(2021)Zhang, Liu, Liu, Wang, and
  Wang]{zhang_super-rational_2021}
Feng Zhang, Yan-Ping Liu, Yan-Ping Liu, Rui-Wu Wang, and Si-Yi Wang.
\newblock Super-rational aspiration induced strategy updating promotes
  cooperation in the asymmetric prisoner's dilemma game.
\newblock \emph{Applied Mathematics and Computation}, 403:\penalty0 126180,
  August 2021.
\newblock \doi{10.1016/J.AMC.2021.126180}.
\newblock URL \url{https://doi.org/10.1016/j.amc.2021.126180}.

\bibitem[Szolnoki et~al.(2011)Szolnoki, Szabó, and Perc]{szolnoki_phase_2011}
Attila Szolnoki, György Szabó, and Matjaž Perc.
\newblock Phase diagrams for the spatial public goods game with pool
  punishment.
\newblock \emph{Physical Review E}, 83\penalty0 (3):\penalty0 036101--036101,
  March 2011.
\newblock \doi{10.1103/PHYSREVE.83.036101}.
\newblock URL
  \url{https://journals.aps.org/pre/abstract/10.1103/PhysRevE.83.036101}.

\bibitem[Perc et~al.(2013)Perc, Gómez-Gardenes, Szolnoki, Floría, and
  Moreno]{perc_evolutionary_2013}
Matjaž Perc, Jesús Gómez-Gardenes, Attila Szolnoki, Luis~M Floría, and
  Yamir Moreno.
\newblock Evolutionary dynamics of group interactions on structured
  populations: a review.
\newblock \emph{Journal of the Royal Society Interface}, 10\penalty0
  (80):\penalty0 20120997--20120997, March 2013.
\newblock \doi{10.1098/RSIF.2012.0997}.
\newblock URL \url{https://doi.org/10.1098/rsif.2012.0997}.

\bibitem[Hauser et~al.(2019)Hauser, Hilbe, Chatterjee, and
  Nowak]{hauser_social_2019}
Oliver~P. Hauser, Christian Hilbe, Krishnendu Chatterjee, and Martin~A. Nowak.
\newblock Social dilemmas among unequals.
\newblock \emph{Nature}, 572\penalty0 (7770):\penalty0 524--527, August 2019.
\newblock \doi{10.1038/S41586-019-1488-5}.
\newblock URL \url{https://www.nature.com/articles/s41586-019-1488-5}.

\bibitem[Latora et~al.(2021)Latora, Latora, Latora, Moreno, Moreno, Arruda,
  Battiston, Battiston, Perc, Perc, and
  Alvarez-Rodriguez]{latora_evolutionary_2021}
Vito Latora, Vito Latora, Vito Latora, Yamir Moreno, Yamir Moreno, Guilherme
  Ferraz~de Arruda, Federico Battiston, Federico Battiston, Matjaž Perc,
  Matjaž Perc, and Unai Alvarez-Rodriguez.
\newblock Evolutionary dynamics of higher-order interactions in social
  networks.
\newblock \emph{Nature Human Behaviour}, 5\penalty0 (5):\penalty0 586--595,
  January 2021.
\newblock \doi{10.1038/S41562-020-01024-1}.
\newblock URL \url{https://www.nature.com/articles/s41562-020-01024-1}.

\bibitem[Wang and Szolnoki(2022)]{WANG2022127307}
Chaoqian Wang and Attila Szolnoki.
\newblock Involution game with spatio-temporal heterogeneity of social
  resources.
\newblock \emph{Applied Mathematics and Computation}, 430:\penalty0 127307,
  2022.
\newblock ISSN 0096-3003.
\newblock \doi{https://doi.org/10.1016/j.amc.2022.127307}.
\newblock URL
  \url{https://www.sciencedirect.com/science/article/pii/S0096300322003812}.

\bibitem[Li(2023)]{li_involution_2023}
Bo~Li.
\newblock Involution game with specialization strategy.
\newblock \emph{Advances in Complex Systems}, 26\penalty0 (07n08):\penalty0
  2350013, 2023.
\newblock URL \url{https://doi.org/10.1142/S0219525923500133}.

\bibitem[Huang and Wang(2024)]{huang_memory_based_2024}
Chaochao Huang and Chaoqian Wang.
\newblock Memory-based involution dilemma on square lattices.
\newblock \emph{Chaos, Solitons \& Fractals}, 178:\penalty0 114384--114384,
  January 2024.
\newblock URL
  \url{https://www.sciencedirect.com/science/article/pii/S0960077923012869}.

\bibitem[Chaocheng et~al.(2023)Chaocheng, Qian, Xinru, Renxian, Fuzhen, Yuchi,
  and Jiang]{chaocheng_involution-cooperation-lying_2023}
He~Chaocheng, Huang Qian, Li~Xinru, Zuo Renxian, Liu Fuzhen, Wei Yuchi, and
  Wu~Jiang.
\newblock Involution-cooperation-lying flat game on a network-structured
  population in the group competition.
\newblock \emph{IEEE Transactions on Computational Social Systems}, July 2023.
\newblock \doi{10.1109/TCSS.2023.3287772}.
\newblock URL \url{https://ieeexplore.ieee.org/document/10177703}.

\bibitem[Szolnoki et~al.(2009{\natexlab{b}})Szolnoki, Perc, and
  Szabó]{szolnoki_topology-independent_2009}
Attila Szolnoki, Matjaž Perc, and György Szabó.
\newblock Topology-independent impact of noise on cooperation in spatial public
  goods games.
\newblock \emph{Physical Review E}, 80\penalty0 (5):\penalty0 056109, November
  2009{\natexlab{b}}.
\newblock \doi{10.1103/PHYSREVE.80.056109}.
\newblock URL
  \url{https://journals.aps.org/pre/abstract/10.1103/PhysRevE.80.056109}.

\bibitem[Helbing and Yu(2009)]{doi:10.1073/pnas.0811503106}
Dirk Helbing and Wenjian Yu.
\newblock The outbreak of cooperation among success-driven individuals under
  noisy conditions.
\newblock \emph{Proceedings of the National Academy of Sciences}, 106\penalty0
  (10):\penalty0 3680--3685, 2009.
\newblock \doi{10.1073/pnas.0811503106}.
\newblock URL \url{https://www.pnas.org/doi/abs/10.1073/pnas.0811503106}.

\bibitem[Chen et~al.(2012)Chen, Szolnoki, and Perc]{chen_risk-driven_2012}
Xiaojie Chen, Attila Szolnoki, and Matjaž Perc.
\newblock Risk-driven migration and the collective-risk social dilemma.
\newblock \emph{Physical Review E}, 86\penalty0 (3):\penalty0 036101, September
  2012.
\newblock \doi{10.1103/PHYSREVE.86.036101}.
\newblock URL
  \url{https://journals.aps.org/pre/abstract/10.1103/PhysRevE.86.036101}.

\bibitem[Wang and Huang(2022)]{WANG2022128097}
Chaoqian Wang and Chaochao Huang.
\newblock Between local and global strategy updating in public goods game.
\newblock \emph{Physica A: Statistical Mechanics and its Applications},
  606:\penalty0 128097, 2022.
\newblock ISSN 0378-4371.
\newblock \doi{https://doi.org/10.1016/j.physa.2022.128097}.
\newblock URL
  \url{https://www.sciencedirect.com/science/article/pii/S0378437122006793}.

\bibitem[Zhang et~al.(2022)Zhang, Li, Dai, and Yang]{ZHANG2022127073}
Liming Zhang, Haihong Li, Qionglin Dai, and Junzhong Yang.
\newblock Migration based on environment comparison promotes cooperation in
  evolutionary games.
\newblock \emph{Physica A: Statistical Mechanics and its Applications},
  595:\penalty0 127073, 2022.
\newblock ISSN 0378-4371.
\newblock \doi{https://doi.org/10.1016/j.physa.2022.127073}.
\newblock URL
  \url{https://www.sciencedirect.com/science/article/pii/S0378437122001224}.

\bibitem[Li et~al.(2021)Li, Dai, Yang, Zhang, and Huang]{li_effects_2021}
Haihong Li, Qionglin Dai, Junzhong Yang, Lan Zhang, and Changwei Huang.
\newblock Effects of directional migration for pursuit of profitable
  circumstances in evolutionary games.
\newblock \emph{Chaos Solitons \& Fractals}, 144:\penalty0 110709, March 2021.
\newblock \doi{10.1016/J.CHAOS.2021.110709}.
\newblock URL
  \url{https://www.sciencedirect.com/science/article/pii/S096007792100062X}.

\bibitem[Shaw et~al.(2023)Shaw, Torstenson, Craft, and Binning]{shaw_gaps_2023}
Allison~K Shaw, Martha Torstenson, Meggan~E Craft, and Sandra~A Binning.
\newblock Gaps in modelling animal migration with evolutionary game theory:
  infection can favour the loss of migration.
\newblock \emph{Philosophical Transactions of the Royal Society B},
  378\penalty0 (1876):\penalty0 20210506--20210506, May 2023.
\newblock \doi{10.1098/RSTB.2021.0506}.
\newblock URL \url{https://doi.org/10.1098/rstb.2021.0506}.

\bibitem[Buesser et~al.(2013)Buesser, Tomassini, and
  Antonioni]{buesser_opportunistic_2013}
Pierre Buesser, Marco Tomassini, and Alberto Antonioni.
\newblock Opportunistic migration in spatial evolutionary games.
\newblock \emph{Physical Review E}, 88\penalty0 (4):\penalty0 042806, October
  2013.
\newblock \doi{10.1103/PHYSREVE.88.042806}.
\newblock URL
  \url{https://journals.aps.org/pre/abstract/10.1103/PhysRevE.88.042806}.

\bibitem[Xiao et~al.(2022)Xiao, Zhang, Li, Dai, and
  Yang]{xiao_environment-driven_2022}
Shilin Xiao, Liming Zhang, Haihong Li, Qionglin Dai, and Junzhong Yang.
\newblock Environment-driven migration enhances cooperation in evolutionary
  public goods games.
\newblock \emph{The European Physical Journal B}, 95\penalty0 (4):\penalty0 67,
  2022.
\newblock URL
  \url{https://link.springer.com/article/10.1140/epjb/s10051-022-00327-8}.

\bibitem[Li and Ye(2015)]{li_effect_2015}
Yan Li and Hang Ye.
\newblock Effect of migration based on strategy and cost on the evolution of
  cooperation.
\newblock \emph{Chaos, Solitons \& Fractals}, 76:\penalty0 156--165, July 2015.
\newblock \doi{10.1016/J.CHAOS.2015.04.006}.
\newblock URL
  \url{https://www.sciencedirect.com/science/article/pii/S0960077915001174}.

\bibitem[Dhakal et~al.(2022)Dhakal, Chiong, Chica, and
  Han]{dhakal_evolution_2022}
Sandeep Dhakal, Raymond Chiong, Manuel Chica, and The~Anh Han.
\newblock Evolution of cooperation and trust in an {N}-player social dilemma
  game with tags for migration decisions.
\newblock \emph{Royal Society Open Science}, 9\penalty0 (5):\penalty0
  212000--212000, May 2022.
\newblock URL \url{https://doi.org/10.1098/rsos.212000}.

\bibitem[Sun et~al.(2026)Sun, Feng, Kang, Shen, and Chen]{sun_adaptive_2026}
Xingping Sun, Kaiyu Feng, Hongwei Kang, Yong Shen, and Qingyi Chen.
\newblock Adaptive migration guided by reputation-pheromone dynamics in spatial
  games.
\newblock \emph{Chaos, Solitons \& Fractals}, 202:\penalty0 117440--117440,
  January 2026.
\newblock URL
  \url{https://www.sciencedirect.com/science/article/pii/S0960077925014535}.

\bibitem[Szabó et~al.(2005)Szabó, Vukov, and Szolnoki]{szabo2005phase}
György Szabó, Jeromos Vukov, and Attila Szolnoki.
\newblock Phase diagrams for an evolutionary prisoner's dilemma game on
  two-dimensional lattices.
\newblock \emph{Physical Review E}, 72\penalty0 (4):\penalty0 047107, October
  2005.
\newblock \doi{10.1103/PHYSREVE.72.047107}.
\newblock URL
  \url{https://journals.aps.org/pre/abstract/10.1103/PhysRevE.72.047107}.

\bibitem[Vukov et~al.(2006)Vukov, Szabó, and Szolnoki]{Vukov20026}
Jeromos Vukov, György Szabó, and Attila Szolnoki.
\newblock Cooperation in the noisy case: {Prisoner}'s dilemma game on two types
  of regular random graphs.
\newblock \emph{Physical Review E}, 73\penalty0 (6):\penalty0 067103, June
  2006.
\newblock \doi{10.1103/PHYSREVE.73.067103}.
\newblock URL
  \url{https://journals.aps.org/pre/abstract/10.1103/PhysRevE.73.067103}.

\bibitem[Achdou et~al.(2020)Achdou, Cardaliaguet, Delarue, Porretta,
  Santambrogio, Cardaliaguet, and Porretta]{achdou_introduction_2020}
Yves Achdou, Pierre Cardaliaguet, François Delarue, Alessio Porretta, Filippo
  Santambrogio, Pierre Cardaliaguet, and Alessio Porretta.
\newblock An {Introduction} to {Mean} {Field} {Game} {Theory}.
\newblock pages 1--158, January 2020.
\newblock \doi{10.1007/978-3-030-59837-2_1}.
\newblock URL
  \url{https://link.springer.com/chapter/10.1007/978-3-030-59837-2_1}.

\bibitem[Sandholm(2020)]{sandholm_evolutionary_2020}
William~H Sandholm.
\newblock \emph{Evolutionary game theory}.
\newblock Springer, 2020.
\newblock URL
  \url{https://link.springer.com/rwe/10.1007/978-1-0716-0368-0_188}.

\bibitem[Antonov et~al.(2021)Antonov, Burovski, and
  Shchur]{antonov_mean-field_2021}
Dmitriy Antonov, Evgeni Burovski, and Lev Shchur.
\newblock Mean-field interactions in evolutionary spatial games.
\newblock \emph{Physical Review Research}, 3\penalty0 (3):\penalty0
  L032072--L032072, September 2021.
\newblock URL
  \url{https://journals.aps.org/prresearch/abstract/10.1103/PhysRevResearch.3.L032072}.

\bibitem[Traulsen et~al.(2006)Traulsen, Nowak, and
  Pacheco]{traulsen_stochastic_2006}
Arne Traulsen, Martin~A. Nowak, and Jorge~M. Pacheco.
\newblock Stochastic dynamics of invasion and fixation.
\newblock \emph{Physical Review E}, 74\penalty0 (1):\penalty0 011909--011909,
  July 2006.
\newblock \doi{10.1103/PHYSREVE.74.011909}.
\newblock URL
  \url{https://journals.aps.org/pre/abstract/10.1103/PhysRevE.74.011909}.

\bibitem[Schlag(1998)]{schlag_why_1998}
Karl~H Schlag.
\newblock Why imitate, and if so, how?: {A} boundedly rational approach to
  multi-armed bandits.
\newblock \emph{Journal of economic theory}, 78\penalty0 (1):\penalty0
  130--156, 1998.
\newblock URL
  \url{https://www.sciencedirect.com/science/article/pii/S0022053197923474}.

\bibitem[Sigmund and Hofbauer(1998)]{sigmund_evolutionary_1998}
Karl Sigmund and Josef Hofbauer.
\newblock \emph{Evolutionary games and population dynamics}.
\newblock Cambridge University Press, January 1998.
\newblock URL
  \url{https://www.cambridge.org/cn/universitypress/subjects/mathematics/mathematical-biology/evolutionary-games-and-population-dynamics}.

\bibitem[Hauert and Szabó(2005)]{hauert_game_2005}
Christoph Hauert and György Szabó.
\newblock Game theory and physics.
\newblock \emph{American Journal of Physics}, 73\penalty0 (5):\penalty0
  405--414, May 2005.
\newblock \doi{10.1119/1.1848514}.
\newblock URL \url{https://doi.org/10.1119/1.1848514}.

\bibitem[Nowak(2006)]{nowak_evolutionary_2006}
Martin~A. Nowak.
\newblock \emph{Evolutionary {Dynamics}: {Exploring} the {Equations} of
  {Life}}.
\newblock Harvard University Press, September 2006.
\newblock URL \url{https://www.jstor.org/stable/j.ctvjghw98}.

\bibitem[Roca and Helbing(2011)]{roca_emergence_2011}
Carlos~P Roca and Dirk Helbing.
\newblock Emergence of social cohesion in a model society of greedy, mobile
  individuals.
\newblock \emph{Proceedings of the National Academy of Sciences of the United
  States of America}, 108\penalty0 (28):\penalty0 11370--11374, July 2011.
\newblock \doi{10.1073/PNAS.1101044108}.
\newblock URL \url{https://www.pnas.org/doi/10.1073/pnas.1101044108}.

\bibitem[Sadhukhan et~al.(2021{\natexlab{a}})Sadhukhan, Chattopadhyay, and
  Chakraborty]{sadhukhan_cooperators_2021}
Shubhadeep Sadhukhan, Rohitashwa Chattopadhyay, and Sagar Chakraborty.
\newblock Cooperators overcome migration dilemma through synchronization.
\newblock \emph{Physical Review Research}, 3\penalty0 (1):\penalty0 013009,
  January 2021{\natexlab{a}}.
\newblock \doi{10.1103/PHYSREVRESEARCH.3.013009}.
\newblock URL
  \url{https://journals.aps.org/prresearch/abstract/10.1103/PhysRevResearch.3.013009}.

\bibitem[Sadhukhan et~al.(2021{\natexlab{b}})Sadhukhan, Chattopadhyay, and
  Chakraborty]{sadhukhan_amplitude_2021}
Shubhadeep Sadhukhan, Rohitashwa Chattopadhyay, and Sagar Chakraborty.
\newblock Amplitude death in coupled replicator map lattice: {Averting}
  migration dilemma.
\newblock \emph{Physical Review E}, 104\penalty0 (4):\penalty0 044304--044304,
  October 2021{\natexlab{b}}.
\newblock URL
  \url{https://journals.aps.org/pre/abstract/10.1103/PhysRevE.104.044304}.

\bibitem[Lee et~al.(2022)Lee, Cleveland, and Szolnoki]{10.1063/5.0100772}
Hsuan-Wei Lee, Colin Cleveland, and Attila Szolnoki.
\newblock When costly migration helps to improve cooperation.
\newblock \emph{Chaos: An Interdisciplinary Journal of Nonlinear Science},
  32\penalty0 (9):\penalty0 093103, 09 2022.
\newblock ISSN 1054-1500.
\newblock \doi{10.1063/5.0100772}.
\newblock URL \url{https://doi.org/10.1063/5.0100772}.

\end{thebibliography}






\end{document}